%% file: main.tex
\documentclass[manuscript]{acmart}

\AtBeginDocument{%
  }

\setcopyright{acmlicensed}
\copyrightyear{2026}
\acmYear{2026}
\acmDOI{XXXXXXX.XXXXXXX}

\acmJournal{TOSEM}

\usepackage{ulem}  
\usepackage{graphicx}
\usepackage{multirow}
\usepackage[title]{appendix}
\usepackage{xcolor}
\usepackage{manyfoot}
\usepackage{algorithm}
\usepackage{algorithmicx}
\usepackage{algpseudocode}
\usepackage{booktabs}
\usepackage{multirow}
\usepackage{listings}
\usepackage{bookmark}
\usepackage{amsmath}
\usepackage{pgfplots}
\usepgfplotslibrary{groupplots}
\pgfplotsset{compat=1.18}
\graphicspath{{../assets/}}

\lstdefinelanguage{Rust}{
  keywords={fn, let, mut, const, static, struct, enum, impl, trait, pub, priv, unsafe, async, await, move, where, use, mod, crate, self, super, as, if, else, match, while, for, loop, break, continue, return, type, dyn, Box, Option, Result, String, Vec},
  keywordstyle=\color{blue}\bfseries,
  ndkeywords={Self, None, Some, Ok, Err, true, false},
  ndkeywordstyle=\color{darkgray}\bfseries,
  sensitive=true,
  comment=[l]//,
  morecomment=[s]{/*}{*/},
  commentstyle=\color{gray}\ttfamily,
  stringstyle=\color{red}\ttfamily,
  morestring=[b]",
  morestring=[b]',
}

\theoremstyle{plain}

\theoremstyle{definition}

\begin{document}

\title{AdaTrans: Automated C-to-Rust Transformation via Error-Adaptive Repair}

\author{Xiaofan Liu}
\email{xfliu@whu.edu.cn}
\affiliation{%
  \institution{School of Computer Science}%
  \institution{Wuhan University}%
  \country{China}
  \city{Wuhan}
}

\author{Zhuang Zhao}
\email{zhao2z@whu.edu.cn}
\affiliation{%
  \institution{School of Computer Science}%
  \institution{Wuhan University}%
  \country{China}
  \city{Wuhan}
}

\author{Zecan Li}
\email{zecanli@whu.edu.cn}
\affiliation{%
  \institution{School of Computer Science}%
  \institution{Wuhan University}%
  \country{China}
  \city{Wuhan}
}

\author{Ziqi Shuai}
\email{szq@whu.edu.cn}
\affiliation{%
  \institution{School of Computer Science}%
  \institution{Wuhan University}%
  \country{China}
  \city{Wuhan}
}

\author{Yanming Yang}
\email{yanming.yang@whu.edu.cn}
\affiliation{%
  \institution{School of Computer Science}%
  \institution{Wuhan University}%
  \country{China}
  \city{Wuhan}
}

\author{Qi Xin}
\email{qxin@whu.edu.cn}
\affiliation{%
  \institution{School of Computer Science}%
  \institution{Wuhan University}%
  \country{China}
  \city{Wuhan}
}

\author{Jifeng Xuan}
\authornote{Corresponding author.}
\email{jxuan@whu.edu.cn}
\affiliation{%
  \institution{School of Computer Science}%
  \institution{Wuhan University}%
  \country{China}
  \city{Wuhan}
}

\input{sections/abstract}



\maketitle
\raggedbottom
\input{sections/introduction}
\input{sections/background}
\input{sections/methodology}
\input{sections/experimental-setup}
\input{sections/experimental-evaluation}
\input{sections/threats}
\input{sections/related-work}

\input{sections/conclusion}
\input{sections/declarations}
\input{sections/acknowledgments}

\bibliographystyle{ACM-Reference-Format}
\bibliography{references}

\end{document}

%% file: sections/abstract.tex
\begin{abstract}
    Automated transformation of C to Rust remains challenging due to the strict ownership system and borrowing semantics of Rust.
    Large language models (LLMs) show potential for code generation, but frequently produce Rust code that violates ownership rules or relies on \texttt{unsafe} blocks. Three factors compound this problem. 
    First, the ownership system of Rust presents a semantic barrier that general-purpose LLMs cannot reliably cross. Second, existing approaches underutilize compiler feedback for program repair.
    Third, without targeted verification, generated code tends to circumvent safety constraints rather than satisfy them.
    We propose AdaTrans, a framework that addresses these challenges by integrating three mechanisms.
    First, a Strategy-Driven Retrieval-Augmented Generation (RAG) mechanism maps compiler errors to specific repair strategies.
    Second, an Error-Stratified Transformation Strategy (ESTS) classifies compiler diagnostics into semantic error categories. 
    Per-category temperature scheduling and stagnation detection adapt the repair behavior to balance generation diversity with constraint satisfaction.
    Third, a multi-stage validation pipeline ensures both compilability and functional equivalence through iterative repair.
    We focus on file-level transformation of self-contained C modules with standard input/output behavior, a controlled setting that isolates the core semantic mapping from C to Rust and admits differential-testing oracles for rigorous functional-equivalence checking.
    We evaluate AdaTrans on a dataset of 104 algorithmic problems from LeetCode Weekly Contests. 
    We compare it with three existing LLM-based C-to-Rust tools, a zero-shot LLM baseline, and the c2rust AST-level transpiler.
    Across three independent runs, AdaTrans achieves a mean compilation pass rate of 95.51\% ($\pm$ 1.11\%) and a mean solve rate of 81.09\% ($\pm$ 3.09\%)  under a fuzz-based test oracle, with a mean \texttt{unsafe} file rate of 1.19\%. 
    AdaTrans improves the solve rate over the strongest existing LLM-based tool by 59.94 percentage points.
    It also keeps the \texttt{unsafe} file rate at 1.19\%, well below that of the c2rust AST-level transpiler.
    These results demonstrate that adapting repair strategies to the semantic characteristics of compiler diagnostics can reconcile transformation correctness with memory safety.
\end{abstract}

%% file: sections/introduction.tex
\section{Introduction}\label{sec:intro}
System programming languages underpin operating systems, embedded devices, and high-performance computing infrastructure.
Among these languages, C has maintained its position as the de facto standard for over five decades. This enduring dominance stems from its low-level hardware control and high execution efficiency.
However, the unrestrained memory access capabilities of C represent a critical weakness, which manifests as serious security vulnerabilities, including buffer overflows, null pointer dereferences, and data races.
The prevalence of memory-related vulnerabilities in system software motivates the search for safer programming language alternatives.
Rust has emerged as a prominent alternative to C, with widespread adoption in both academia and industry.
The memory safety of Rust rests on three mechanisms: ownership, borrowing, and lifetimes. These mechanisms enforce safety checks at compile time and eliminate the majority of memory errors without sacrificing runtime performance.

The transformation of codebases from C to Rust remains a challenging engineering task that traditionally demands extensive manual effort.
Existing automated tools, such as c2rust~\cite{c2rust} and Laertes~\cite{laet}, generate syntactically correct Rust code that relies heavily on \texttt{unsafe} blocks to bypass the borrow checker.
Such reliance on \texttt{unsafe} blocks effectively defeats the original purpose of memory safety during migration to Rust~\cite{unsafe_migration}.
Large Language Models (LLMs), pre-trained on massive open-source codebases, have recently demonstrated potential in code generation and transformation~\cite{llm_code_survey,codex} by capturing complex syntactic structures across multiple programming languages.
General-purpose LLMs, however, still face severe challenges in C-to-Rust transformation.
First, the ownership rules of Rust impose a semantic barrier that general-purpose models cannot reliably cross. 
Generated code frequently violates borrowing and lifetime invariants because the LLM fails to map manual memory management in C to the explicit ownership model of Rust.
Second, existing approaches underutilize compiler feedback for iterative repair. 
Rust compiler diagnostics encode rich semantic information, yet LLMs struggle to interpret complex error messages, which restricts automated repairs to superficial syntactic adjustments rather than structural corrections.
Third, without targeted verification mechanisms, most generation methods prioritize syntactic correctness over functional integrity. \texttt{Unsafe} blocks and logical errors therefore persist in the generated code.

Existing research on C-to-Rust transformation spans multiple granularities.
Recent LLM-based approaches~\cite{evotrans,tymcrat,ptrtrans} primarily target project-level transformation, improving code idiomaticity through techniques such as skeleton-guided evolution, type-directed mutation, and pointer knowledge graphs.
These advances are complementary to our work, because the central difficulty of safe transformation persists across granularities. 
It lies in the semantic mapping between manual memory management in C and the ownership model of Rust, where
functional equivalence and memory safety must hold simultaneously.
We isolate this semantic core by studying file-level transformation of self-contained C modules with standard input/output behavior. 
This scope is a deliberate methodological choice rather than a capability limitation.
Self-contained modules admit ground-truth differential-testing oracles, making functional equivalence directly verifiable, whereas project-level migration rarely provides comparable validation mechanisms. 
Our repair mechanism treats each file-unit as a minimal project, so it is not bound to this granularity.
The controlled setting removes confounding factors such as build configuration, cross-module dependencies, and external bindings, allowing us to focus on semantic fidelity and safety.
Such units also arise naturally during incremental migration of legacy C codebases~\cite{rustforlinux}, where standalone utilities and self-contained libraries are often migrated first before larger project-wide
integration, while remaining straightforward to validate in isolation~\cite{ctorust_userstudy}.

We assess the current state of file-level C-to-Rust transformation under this controlled setting. 
For this purpose, we construct a dataset of 104 algorithmic problems sourced from LeetCode Weekly Contests released after the knowledge cutoff of the LLM snapshot used in our experiments (Section~\ref{sec:setup}).
This design mitigates, but does not eliminate, the risk of benchmark contamination with respect to the evaluated model. 
On this dataset, a zero-shot LLM (\texttt{gpt-4o-mini}) reaches a pass@100 estimate of 70.58\%. 
We compute this estimate from 200 independent samples per problem with the unbiased estimator of Chen et al.~\cite{codex}.
We use this high-budget resampling result as a strong brute-force reference under the same backbone. This reference shows that repeated independent sampling alone is still insufficient for reliable transformation. 
We therefore introduce AdaTrans, an end-to-end framework that transforms individual C files into predominantly safe and compilable Rust programs and preserves behavioral consistency.

The key design intuition of AdaTrans is to adapt repair behavior to the type of failure signal observed during validation. Different error categories correspond to different repair needs, and a uniform strategy across heterogeneous failures is often ineffective. 
Syntax errors, as relatively rigid defects with narrow repair spaces, tend to benefit from low-temperature exploitation. 
Ownership-related errors often require moderate-temperature exploration over alternative memory-management patterns. 
Behavioral failures, where the program compiles but produces incorrect outputs, typically require higher-temperature exploration to escape locally consistent but semantically incorrect solutions~\cite{self_consistency,temp_sensitivity}. 
AdaTrans operationalizes this intuition through an error-driven repair loop that converts validation feedback into signals for retrieval and repair control (Section~\ref{sec:method}).

The main contributions of this paper are as follows:

\begin{enumerate}
    \item \textbf{A compiler-feedback-driven framework for C-to-Rust transformation}. 
    We propose AdaTrans, a three-phase generate-verify-repair pipeline designed for self-contained C files in a controlled file-level migration setting.
    \item \textbf{An operational error-stratified repair strategy}. 
    We introduce ESTS, which groups validation signals into coarse error categories and adapts repair behavior through category-aware temperature scheduling and stagnation recovery.
    \item \textbf{A strategy-driven retrieval mechanism for repair guidance}. 
    We design a RAG module that maps diagnostic signals to repair templates and Rust-specific knowledge, and provides structured guidance beyond raw compiler messages.
    \item \textbf{An empirical study of correctness-safety trade-offs under a fixed LLM backbone}. 
    On a 104-problem dataset designed to mitigate contamination risk with respect to the evaluated model snapshot, AdaTrans achieves a mean compilation pass rate of 95.51\% ($\pm$ 1.11\%) and a mean solve rate of 81.09\% ($\pm$ 3.09\%) across three independent runs, while maintaining a mean \texttt{unsafe} file rate of 1.19\%.
\end{enumerate}

The remainder of this paper is organized as follows. 
Section~\ref{sec:background} introduces the memory model differences between C and Rust, analyzes the limitations of LLM-based code generation, and presents a motivating case study.
Section~\ref{sec:method} details the problem formulation and the three-phase pipeline of the AdaTrans framework.
Section~\ref{sec:setup} describes the experimental setup, and Section~\ref{sec:evaluation} presents the evaluation results.
Section~\ref{sec:threats} discusses the threats to validity.
Section~\ref{sec:related} reviews related work. Finally, Section~\ref{sec:conclusion} concludes the paper.

%% file: sections/background.tex
\section{Background}\label{sec:background}
\subsection{Memory Model Differences between C and Rust}\label{subsec:memorymodel}
The C memory model operates as a direct abstraction of the von Neumann architecture. Programmers manipulate a linear address space through raw pointers. 
This design grants flexibility and performance but places the entire burden of memory safety on the programmer. 
Memory allocated via \texttt{malloc} must be released through \texttt{free} exactly once. The absence of compiler enforcement gives rise to three critical error classes:
\begin{itemize}
    \item \textbf{Use-After-Free (UAF)}: Dereference of a dangling pointer after its underlying allocation has been deallocated.
    \item \textbf{Double Free}: Release of the same allocation more than once, which corrupts heap metadata and enables exploitation.
    \item \textbf{Memory Leak}: Failure to deallocate memory, which causes gradual resource exhaustion.
\end{itemize}

Rust prevents UAF and double-free errors at compile time through an ownership model rooted in an affine type system. 
Automatic resource management (RAII) further mitigates memory leaks by tying deallocation to scope exit. 
For every value $v$, the compiler maintains a unique binding $\mathcal{O}(x, v)$ and enforces three core axioms:
\begin{itemize}
    \item \textbf{Uniqueness}: $\forall v, \forall t: \exists!\, x$ such that $\mathcal{O}(x, v)$ holds at time $t$. Ownership of every resource is unambiguous at every program point.
    \item \textbf{Move Semantics}: An assignment $y = x$ transfers ownership so that $\mathcal{O}(y, v)$ holds and $x$ enters an uninitialized state ($\bot$). This rule prevents the shallow-copy hazards inherent in C pointer aliasing.
    \item \textbf{Borrowing Invariants}: Within any given scope, a value may be borrowed through either multiple immutable references ($\&x$) or exactly one mutable reference (\texttt{\&mut x}), but not both simultaneously. This mutual-exclusion rule eliminates data races by construction.
\end{itemize}
The borrow checker verifies all memory accesses against these axioms through region-based lifetime analysis. 
C-to-Rust transformation therefore demands more than syntactic remapping. 
It requires a structural transformation from unconstrained pointer manipulation to strict ownership enforcement.

\subsection{Limitations of LLM-Based Code Generation}\label{subsec:limits}

LLMs demonstrate strong performance on general programming tasks, yet encounter severe bottlenecks when the target language imposes deterministic static constraints. 
Three architectural conflicts underlie these failures.

\textbf{Stochastic Generation vs.\ Deterministic Constraints.}
LLM code generation follows an autoregressive process that predicts $P(x_t \mid x_{<t})$ from local context. 
This mechanism reproduces high-frequency syntactic patterns effectively but struggles with the non-local nature of Rust ownership. 
Finite context windows impose a myopic bias on synthesis. 
This bias yields code that reads fluently at the local level yet violates global borrowing invariants and lifetime rules.

\textbf{Open-Loop Synthesis and Feedback Deficiency.}
The standard LLM generation pipeline operates as an open-loop system with no integration of compiler feedback. 
In manual development, programmers perform root-cause analysis with the aid of external documentation when they encounter errors. 
Without this iterative feedback loop, LLMs produce syntactically plausible but semantically invalid code. 
Complex diagnostics such as \texttt{borrowed value does not live long enough} remain opaque to general-purpose models. 
The consequence is repetitive, ineffective edits or the introduction of new defects.

\textbf{Unsafe Shortcuts in Generated Code.}
When the borrow checker rejects generated code, LLMs frequently resort to \texttt{unsafe} blocks or raw pointers to suppress compiler errors. 
Qin et al.~\cite{unsafe_rust} document that \texttt{unsafe} usage is widespread in real-world Rust codebases. 
Because system-level training corpora contain abundant unsafe patterns, models learn to reproduce them readily and prioritize syntactic acceptance over semantic safety. 
Such shortcuts silence compilation failures but simultaneously circumvent the safety guarantees that motivate the C-to-Rust migration. 
Effective guidance toward idiomatic safe Rust remains a central challenge for automated transformation.

\subsection{Motivating Case Study}\label{subsec:motivation}
An illustrative case of string collection shows how these limitations manifest in practice. 
In typical C code (Fig.~\ref{lst:c_string_example}), the programmer consolidates multiple strings into a heap-allocated buffer and stores their addresses in a pointer array:

\begin{figure}[htbp]
  \centering
  \includegraphics[width=0.92\columnwidth]{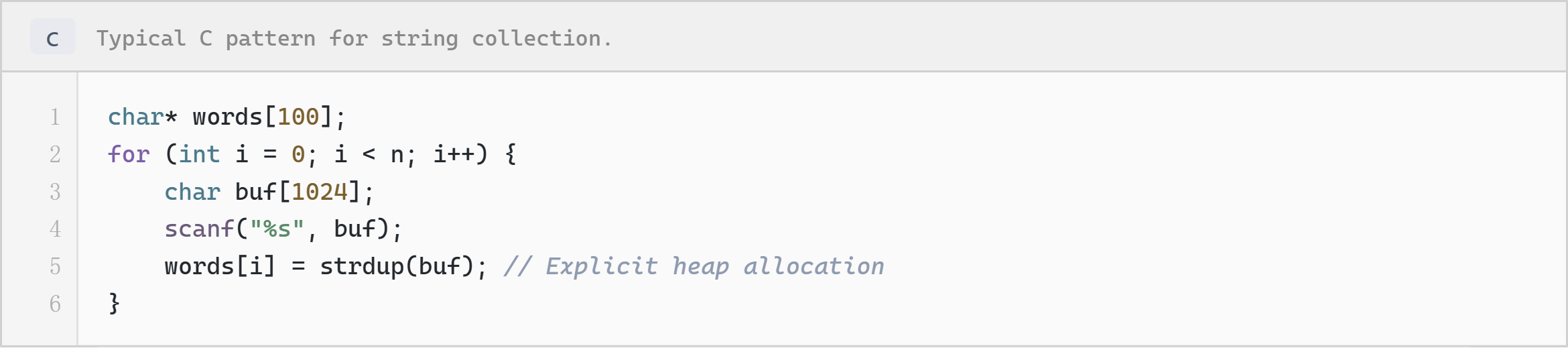}
  \Description{A C code listing that collects strings into a dynamically allocated array of char pointers, illustrating manual memory management through
  malloc and pointer assignment.}
  \caption{Typical C pattern for string collection.}
  \label{lst:c_string_example}
\end{figure}

Direct transformation attempts by general-purpose LLMs typically mimic the pointer-to-buffer logic of the C original, which triggers immediate borrow-checker violations (Fig.~\ref{lst:rust_e0597_example}):

\begin{figure}[htbp]
  \centering
  \includegraphics[width=0.92\columnwidth]{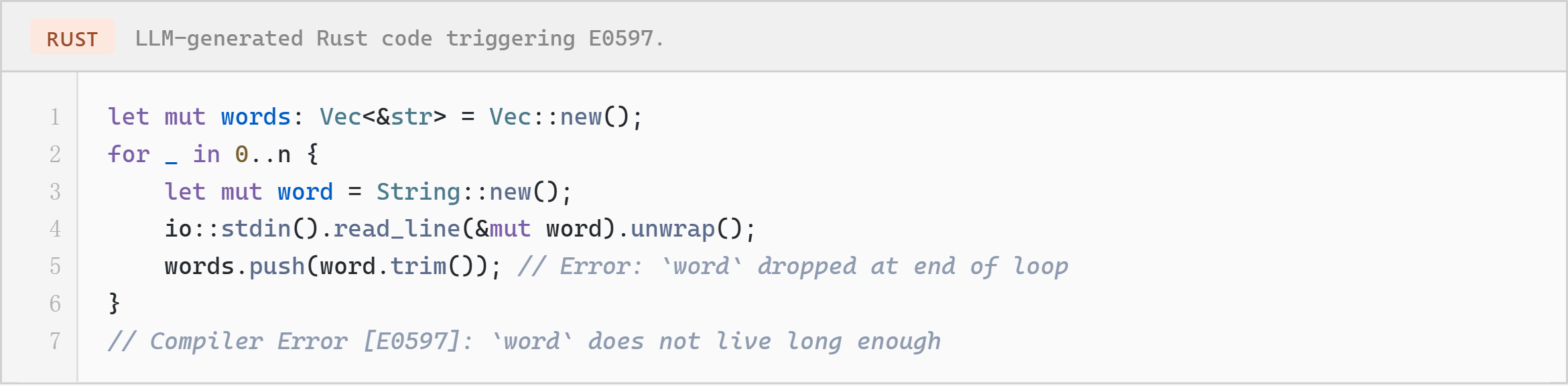}
  \Description{A Rust code listing generated by an LLM that the borrow checker rejects with error E0597, where a borrowed value does not live long
  enough.}
  \caption{LLM-generated Rust code triggering E0597.}
  \label{lst:rust_e0597_example}
\end{figure}

\textbf{Analysis of Model Failure.} General-purpose models struggle to resolve the \texttt{E0597} error for three reasons:

\begin{itemize}
\item \textbf{Superficial Corrections}: 
Models treat compiler errors as syntactic mismatches rather than structural ownership conflicts. 
The result is trial-and-error edits on irrelevant code sections (e.g., substitution of different I/O methods).
\item \textbf{Safety Circumvention}: 
When simple patches fail, models resort to wrapping code in \texttt{unsafe} blocks or calling \texttt{Box::leak}. 
These modifications suppress the error but introduce latent memory leaks, which defeats the purpose of the transformation.
\item \textbf{Insufficient Reasoning Depth}: 
Resolution of this lifetime conflict requires a structural shift from reference storage (\texttt{Vec<\&str>}) to ownership acquisition (\texttt{Vec<String>}). 
Without targeted guidance, models lack the capacity to redesign data structures on the basis of variable lifetimes within iterative scopes.
\end{itemize}


This case illustrates a difficulty that goes beyond syntax: reconciling manual memory management in C with the strict ownership constraints of Rust.
The dataset-wide error analysis in Section~\ref{sec:evaluation} is consistent with this view, where syntactic errors are resolved quickly, ownership-related errors require deeper structural repair, and behavioral errors remain the dominant bottleneck.
AdaTrans addresses these challenges through error-driven adaptation, as described in Section~\ref{sec:method}.

%% file: sections/methodology.tex
\section{Methodology}\label{sec:method}
\subsection{Problem Formulation}\label{subsec:formulation}

We formalize the C-to-Rust transformation as a constrained generation problem. 
Given a source C program $S_C$, the objective is to generate a target Rust program $S_R^*$ that satisfies two strict criteria: compilability (static correctness) and functional equivalence (I/O consistency).

\subsubsection{Functional Equivalence (I/O Consistency)}
Direct proof of semantic equivalence between two programs remains generally undecidable. 
Therefore, we define functional equivalence based on strict Input/Output (I/O) consistency over a representative test suite $\mathcal{I}$. Let $\mathrm{Out}(P, i)$ denote the standard output of program $P$ under test case $i \in \mathcal{I}$. The functional consistency function $\text{SemEquiv}$ is defined as:
\begin{equation}\label{eq:semequiv}
\text{SemEquiv}(S_R, S_C) \iff \forall i \in \mathcal{I}, \text{Norm}(\mathrm{Out}(S_R, i)) = \text{Norm}(\mathrm{Out}(S_C, i))
\end{equation}
where $\text{Norm}(\cdot)$ is a normalization operator that collapses whitespace, trims trailing newlines, and applies floating-point epsilon comparison ($\epsilon \leq 10^{-6}$) when the output domain includes real-valued results. 
In the current evaluation, all problem outputs are exact integers or strings, so only whitespace normalization is applied. The epsilon comparison is retained in the formulation for generality.

\subsubsection{Compilability and Memory Safety}
The target program $S_R$ must pass the strict static analysis of the Rust compiler. Let $\mathcal{V}_{Rust}$ denote the verification process of the compiler, which encompasses type checking, borrow checking, and lifetime analysis. 
The soundness constraint $V(S_R)$ is defined as:
\begin{equation}\label{eq:compilability}
V(S_R) = \text{True} \iff \mathcal{V}_{Rust}(S_R) = \text{Pass}
\end{equation}
This constraint guarantees that the generated code contains no syntax errors and adheres to the ownership rules of Rust, thereby ensuring memory safety for all code outside \texttt{unsafe} blocks.

\subsubsection{Iterative Repair Process}
The generation of $S_R^*$ requires iterative repair to satisfy both criteria. We model this as a sequence of states $\{S_R^{(0)}, S_R^{(1)}, \dots, S_R^{(t)}\}$. 
Each transition relies on the LLM generating a refined version based on previous feedback:
\begin{equation}\label{eq:iterative}
S_R^{(t+1)} \sim P_{LLM}(S_R \mid S_C, E^{(t)}, \mathcal{K}; \theta_t)
\end{equation}
where $E^{(t)}$ represents the diagnostic feedback (compiler errors or test failures) from $S_R^{(t)}$, $\mathcal{K}$ denotes external knowledge retrieved via a Retrieval-Augmented Generation (RAG) module (detailed in Section \ref{subsec:phase2}), and $\theta_t$ is the temperature parameter controlling generation diversity. We use $\theta$ for category-specific temperatures ($\theta_{SL}$, $\theta_{MS}$, $\theta_{LB}$) and $T$ for the base temperature. In Algorithm~\ref{alg:adatrans_main}, the variable $T$ is set by the ESTS mapping $\mathcal{F}_{ESTS}$ and corresponds to $\theta_t$ in Eq.~(\ref{eq:iterative}).

The optimization goal finds the earliest iteration $t$ at which both compilability and I/O consistency are satisfied:
\begin{equation}\label{eq:objective}
t^* = \min\{t \geq 0 \mid V(S_R^{(t)}) \land \text{SemEquiv}(S_R^{(t)}, S_C)\}, \quad S_R^* = S_R^{(t^*)}
\end{equation}
In practice, AdaTrans terminates at the first successful iteration or after exhausting the budget $M$. The iteration count $t^*$ is a termination criterion, not an explicit optimization target.

\subsection{AdaTrans Framework Overview}
The AdaTrans framework addresses the limitations of LLMs in C-to-Rust transformation through an iterative generate-verify-repair paradigm.
This paradigm guides candidate code toward compilation correctness and functional consistency.
AdaTrans iteratively integrates compiler diagnostics and execution feedback from a representative subset of test cases. 
This selective testing strategy verifies core functional requirements without exhausting computational resources during each iteration. 
Based on the test outcomes and compiler error messages, the framework dynamically modulates its transformation strategies. These execution traces, embedded in the model context, provide explicit feedback for the next repair iteration.

\begin{figure}[htbp]
  \centering
  \includegraphics[width=\columnwidth]{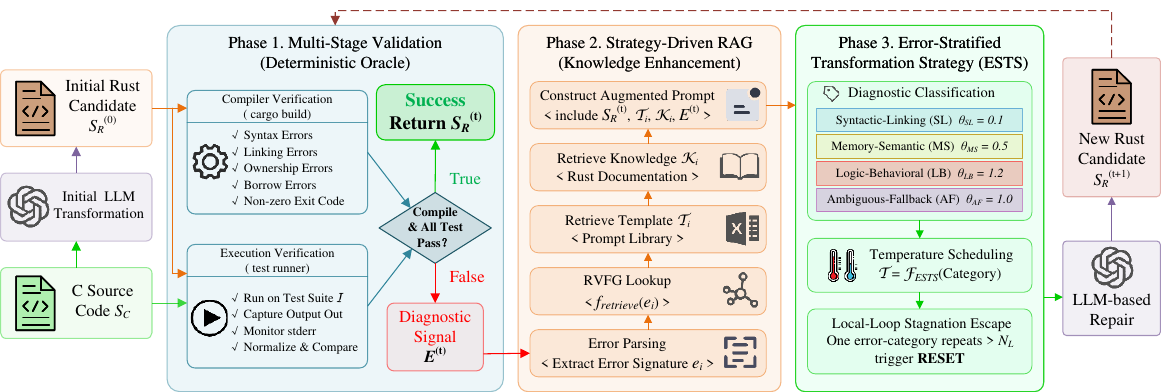}
  \Description{A block diagram of the AdaTrans framework showing three phases, generation, verification, and repair, connected in a closed loop, with the
   Strategy-Driven RAG and ESTS modules feeding the repair phase.}
  \caption{Overall architecture and inter-phase dependencies of the AdaTrans framework.}
  \label{fig:framework}
\end{figure}

Fig.~\ref{fig:framework} illustrates how diagnostic signals flow through the three phases within each repair iteration.
This architecture operates as a closed loop that consists of three sequential phases.
Phase 1 implements a multi-stage validation pipeline to extract diagnostic signals from both the compiler and the execution environment.
Phase 2 employs a Strategy-Driven Retrieval-Augmented Generation (RAG) module to construct adaptive prompt templates based on the diagnostic feedback. This module constructs augmented prompts by retrieving error-specific repair templates and relevant documentation from a Rust knowledge repository.
Phase 3 implements the Error-Stratified Transformation Strategy (ESTS), which classifies diagnostic signals into error categories and adapts the repair behavior accordingly through per-category temperature scheduling and stagnation detection.
The error category extracted during validation drives both the retrieval function in Phase 2 and the temperature scheduling in Phase 3. Syntax errors trigger targeted templates with low-entropy sampling for precise correction, while test failures activate reflection templates with high-entropy sampling to escape local optima.

\subsection{Taxonomy of Violation Signals}\label{subsec:taxonomy}
The complex space $\mathcal{S}_{Rust}$ of valid Rust programs demands a structured search strategy. We establish a taxonomy that classifies the diagnostic feedback $E^{(t)}$ (extracted during the validation phase) according to the \textit{scope of reasoning} and \textit{constraint rigidity} required for repair. This taxonomy partitions the violation signals into four disjoint semantic domains:
\begin{itemize}
    \item \textbf{Syntactic-Linking (SL) Violations}: 
    These represent violations of the \textit{Context-Free Grammar} or basic name resolution (e.g., missing semicolons, unresolved crates). 
    In the state space, SL errors indicate that $S_R$ is not a well-formed string in $\mathcal{L}(Rust)$. 
    These defects are \textbf{deterministic} and typically require only localized, low-entropy sampling for repair.
    \item \textbf{Memory-Semantic (MS) Violations}: 
    These are unique to the Rust affine type system and borrow checker (e.g., \texttt{E0597}, \texttt{E0499}). Unlike SL errors, MS violations often stem from \textit{non-local structural conflicts} between variable lifetimes and ownership transfer. 
    Resolving an MS error requires the LLM to reconstruct the ownership structure. This repair demands moderate-entropy sampling to explore alternative memory management patterns (e.g., cloning vs. borrowing).
    \item \textbf{Logic-Behavioral (LB) Violations}: 
    These occur when $V(S_R) = \text{True}$ but $\text{SemEquiv}(S_R, S_C) = \text{False}$. These signals indicate that while $S_R$ is a valid Rust program, its \textit{execution semantics} deviate from the source. 
    Since the root cause (e.g., algorithmic flaws, off-by-one errors) is often decoupled from the syntax, escaping these local optima requires high-entropy stochastic perturbation to trigger significant structural mutations.
    \item \textbf{Ambiguous-Fallback (AF) Violations}: 
    For unseen error patterns, we define a fallback category for errors not covered by the three categories above, such as those generated by external macros or third-party crates. 
    When an AF violation occurs, the system defaults to baseline-entropy sampling and relies on zero-shot LLM reasoning.
\end{itemize}

This taxonomy directly informs both the retrieval strategy in Phase 2 and the entropy mapping $\mathcal{F}_{ESTS}$ in Phase 3.

\subsection{Phase 1: Multi-Stage Validation Pipeline}
Phase 1 implements a multi-stage validation pipeline to evaluate the generated Rust code. 
This pipeline functions as a deterministic oracle that verifies whether the candidate code satisfies both the strict syntactic invariants of Rust and the functional equivalence of the original C implementation.

Initially, the pipeline invokes the Rust compiler (\texttt{cargo build}) to perform strict static analysis.
This compilation step captures syntax errors, linking failures, and ownership violations. 
The compiler returns non-zero exit codes along with standard error outputs when it detects these structural flaws. 
These outputs contain the exact error codes (e.g., \texttt{E0597}) required for the subsequent repair phases.

If the code compiles successfully, the pipeline proceeds to the dynamic execution stage to evaluate $\text{SemEquiv}(S_R, S_C)$. 
The Test Runner executes the compiled binary over the input domain $\mathcal{I}$ (a predefined fuzzing suite comprising both typical use cases and edge-case mutations) to collect the execution trace $\mathrm{Out}(S_R, i)$. 
The system monitors standard error streams (stderr) during this process to intercept runtime panics (e.g., out-of-bounds access or unhandled \texttt{Result} types) that bypass static analysis.

Output comparison applies the normalization operator $\text{Norm}(\cdot)$ (Equation~\eqref{eq:semequiv}), which collapses whitespace and trims trailing newlines to account for benign formatting differences across system architectures.

\begin{figure}[htbp]
  \centering
  \includegraphics[width=0.92\columnwidth]{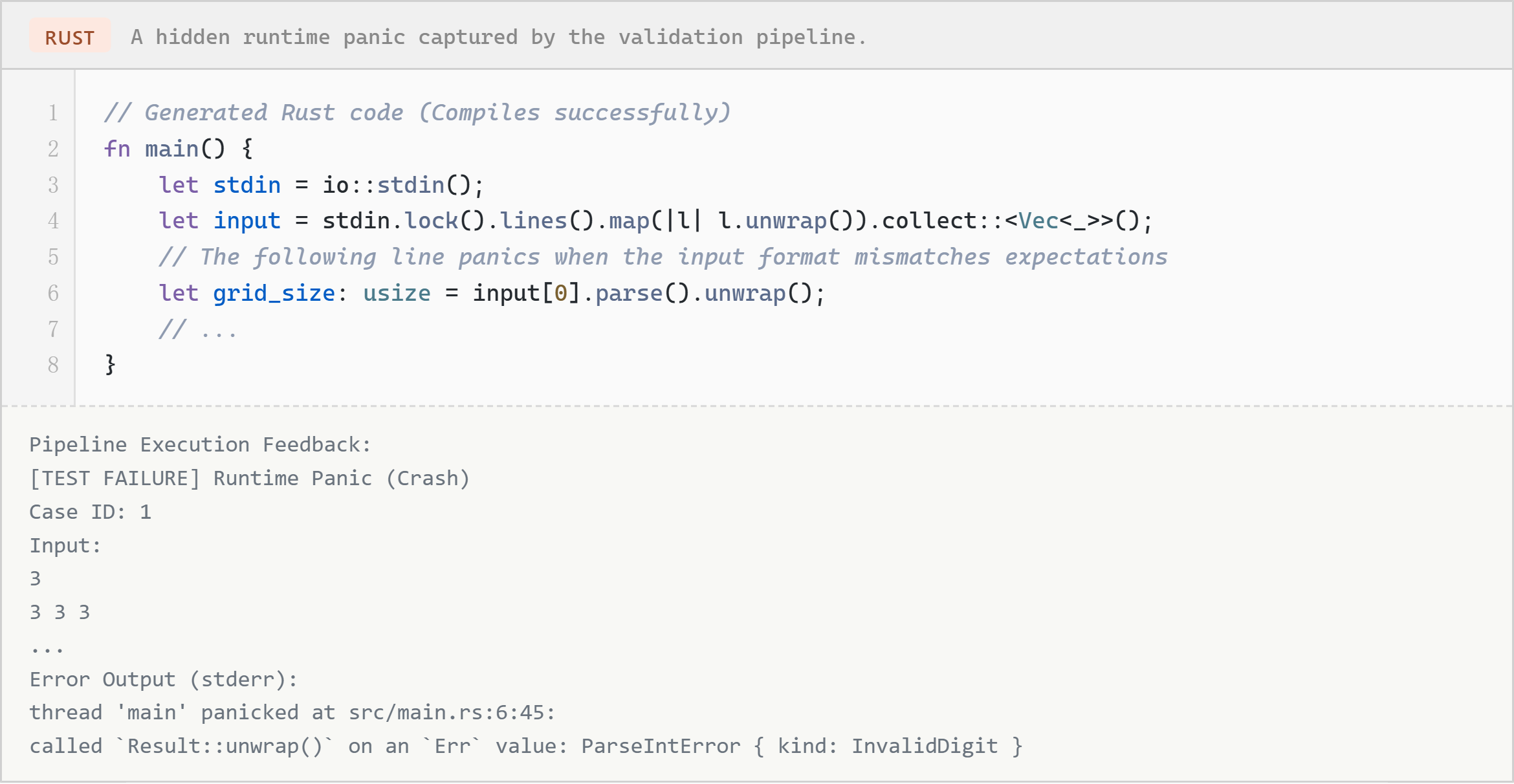}
  \Description{A Rust code listing that compiles successfully but triggers a hidden runtime panic, which the AdaTrans validation pipeline detects during
  test execution.}
  \caption{A hidden runtime panic captured by the validation pipeline during a transformation task.}
  \label{lst:runtime_panic}
\end{figure}

For example, during the evaluation of a dynamic programming transformation task, the generated code passed the strict ownership checks of the compiler but failed the execution stage. 
As shown in Fig.~\ref{lst:runtime_panic}, the validation pipeline successfully caught a hidden panic that bypassed the static compiler checks.

Finally, the validation module categorizes these results into actionable diagnostic signals. 
These signals distinguish between rigid compilation errors (SL, MS) and loose behavioral errors (LB) to provide targeted optimization feedback.

\subsection{Phase 2: Strategy-Driven RAG for Knowledge Enhancement}\label{subsec:phase2}
Phase 2 formalizes a Strategy-Driven Retrieval-Augmented Generation (RAG) mechanism to bridge the gap between raw compiler diagnostics and the structured repair context that LLMs require. 
Recent LLM-based C-to-Rust transformation tools~\cite{evotrans,tymcrat} incorporate compiler feedback into their repair loops, yet they treat diagnostics uniformly without distinguishing error categories. 
Direct injection of raw compiler feedback into LLMs is limited by three factors.
\begin{itemize}
    \item \textbf{Misdiagnosis of Root Causes}: 
    The Rust compiler evaluates ownership strictly at the point of violation, which often misattributes the root cause of a lifetime conflict to a downstream usage rather than its structural origin.
    \item \textbf{Suboptimal Heuristic Suggestions}: 
    The heuristic fixes provided by the compiler (e.g., suggesting \texttt{.clone()} or adding lifetime annotations) prioritize immediate compilation over structural correctness, which misleads the model into adopting suboptimal patterns such as unnecessary cloning or \texttt{Box::leak()}.
    \item \textbf{Absence of Global Context}: 
    Raw feedback lacks the architectural context necessary for global refactoring, which constrains the LLM to localized, often conflicting patches.
\end{itemize}

\begin{figure}[htbp]
  \centering
  \includegraphics[width=\columnwidth]{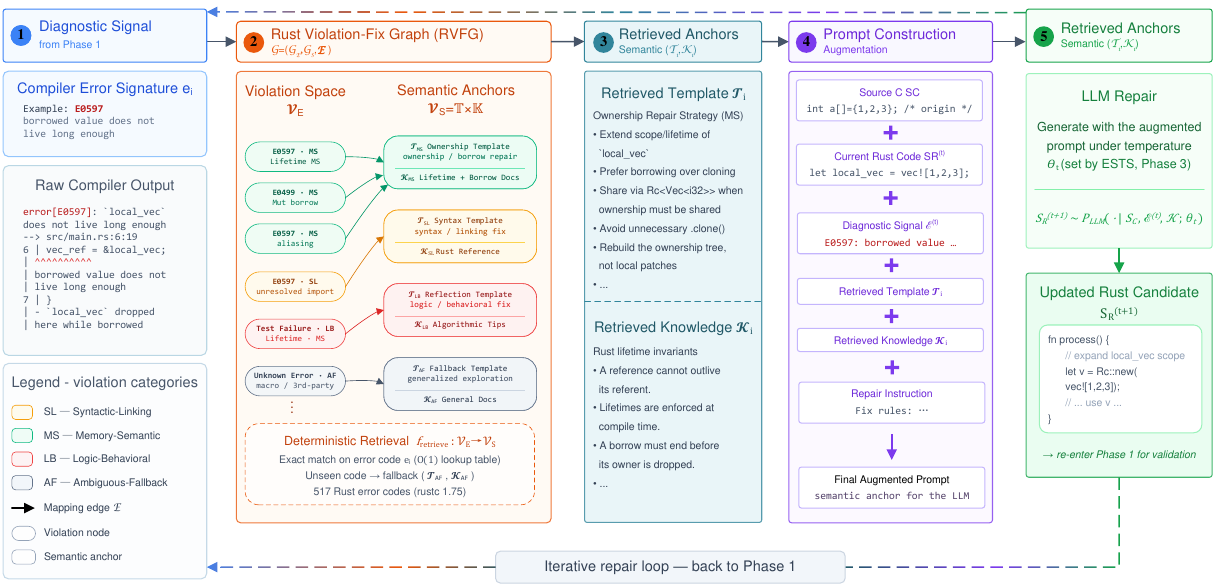}
  \Description{A flow diagram showing how a compiler error signature is mapped through the Rust Violation-Fix Graph to a repair template and official
  Rust documentation tips that are injected into the LLM prompt.}
  \caption{Execution flow of the Strategy-Driven RAG mapping mechanism. The central panel depicts the Rust Violation-Fix Graph $\mathcal{G}=(\mathcal{V}_E,\mathcal{V}_S,\mathcal{E})$, mapping compiler error signatures ($\mathcal{V}_E$) to semantic anchors ($\mathcal{V}_S$) via deterministic edges ($\mathcal{E}$).}
  \label{fig:rag_process}
\end{figure}

We address these limitations through an error-to-template mapping formalized as a \textbf{Rust Violation-Fix Graph (RVFG)}.
We formalize the RVFG as a directed bipartite graph $\mathcal{G} = (\mathcal{V}_E, \mathcal{V}_S, \mathcal{E})$. 
Here, $\mathcal{V}_E$ represents the discrete space of possible Rust compiler error signatures (e.g., \texttt{E0597}), $\mathcal{V}_S = \mathbb{T} \times \mathbb{K}$ represents the set of Semantic Anchors combining repair operators (Prompt Templates $\mathbb{T}$) and official documentation snippets ($\mathbb{K}$), and $\mathcal{E}$ represents the deterministic mapping edges.
As shown in the central component of Fig.~\ref{fig:rag_process}, the violation nodes $\mathcal{V}_E$ (left) are deterministically linked to the semantic anchors $\mathcal{V}_S$ (right) through the mapping edges $\mathcal{E}$.
The Template Library $\mathbb{T}$ comprises a set of role-specific meta-prompts designed for different failure modalities, including a syntax-correction template ($\mathcal{T}_{SL}$), an ownership-reasoning template ($\mathcal{T}_{MS}$), an algorithmic-reflection template ($\mathcal{T}_{LB}$), and a generalized exploration template ($\mathcal{T}_{AF}$).
The RVFG acts as a deterministic mapping from error signatures to repair strategies. 
In practice, this graph reduces to a deterministic lookup table indexed by error code, with $O(1)$ retrieval latency and straightforward extensibility as new error codes are added.
In this graph, nodes represent specific violation semantics defined by exact compiler diagnostic codes (e.g., the 517 distinct Rust compiler error codes extracted from version 1.75), and edges represent the mapping to their corresponding semantic anchors.

We define a deterministic retrieval function $f_{retrieve}: \mathcal{V}_E \rightarrow \mathcal{V}_S$, which is constrained by the edges $\mathcal{E}$ in $\mathcal{G}$. 
When a new diagnostic error $e_i$ is encountered from Phase 1, the module performs exact error-code matching against the nodes in $\mathcal{V}_E$. 
This approach aligns with the principles of deterministic static analysis, as compiler errors are strongly-typed signals whose resolution demands precision rather than semantic approximation. 
If the system encounters an Ambiguous-Fallback (AF) error (e.g., from an unseen third-party macro), $f_{retrieve}$ gracefully degrades to the generalized exploration template $\mathcal{T}_{AF}$ and delegates the repair to zero-shot LLM reasoning. 
The RVFG is not a rigid code-replacement table. Rather, it provides \textit{semantic anchors}. 
The actual code transformation is not hard-coded but dynamically synthesized by the LLM, which applies these semantic constraints to the specific local or inter-procedural context of the target function.

Fig.~\ref{fig:rag_process} illustrates the execution flow of this mapping strategy. 
During the initial step of each repair iteration, the RAG module parses the diagnostic signal from the validation phase. 
The module then evaluates $f_{retrieve}(e_i)$ to construct the augmented prompt.

\begin{figure}[htbp]
  \centering
  \includegraphics[width=0.92\columnwidth]{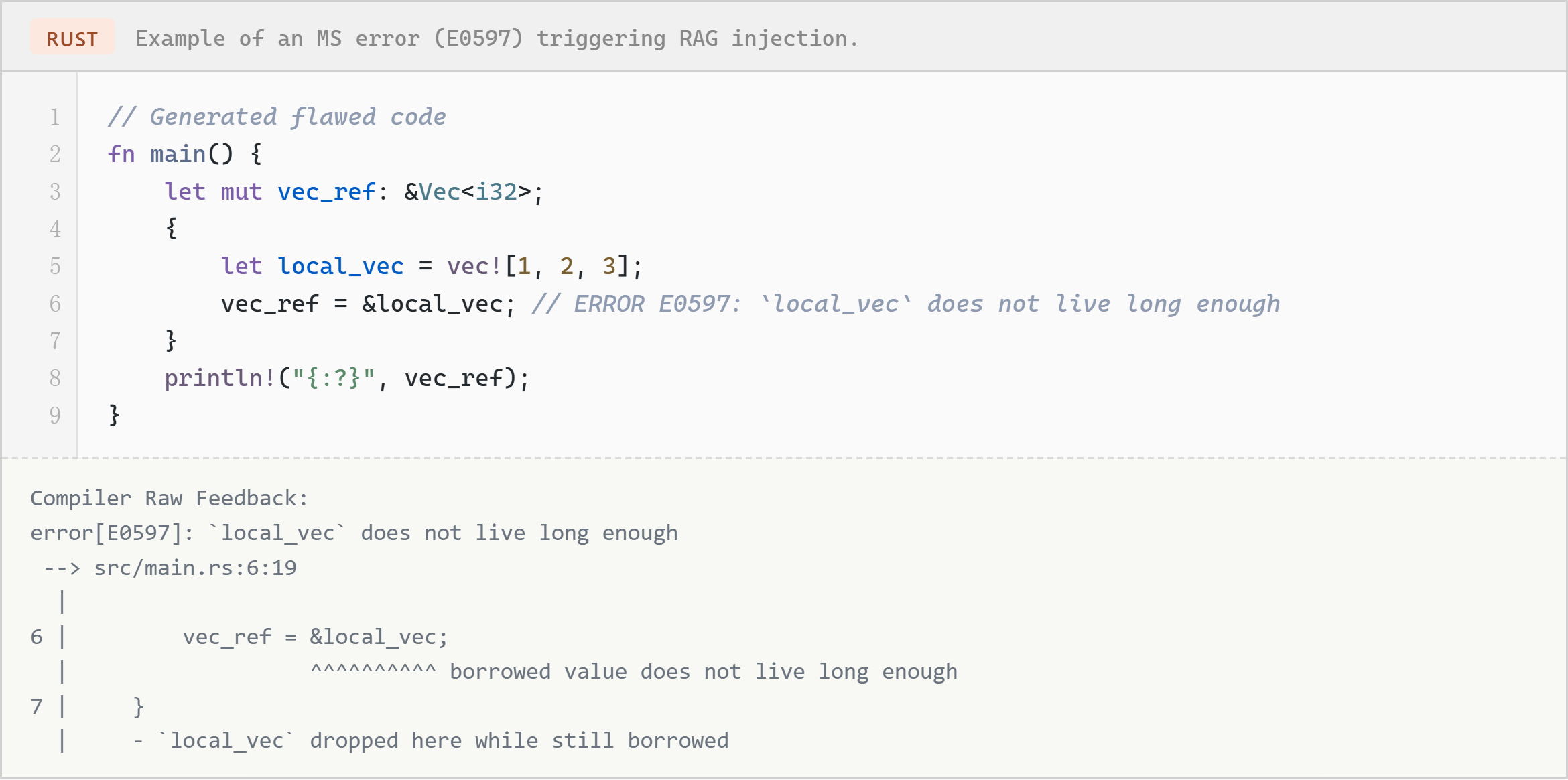}
  \Description{A Rust code listing that produces a Memory-Semantic error E0597, used to illustrate how the matching error signature triggers
  retrieval-augmented repair guidance.}
  \caption{Example of an MS error (E0597) triggering RAG injection.}
  \label{lst:e0597_example}
\end{figure}

For instance, consider Fig.~\ref{lst:e0597_example}, which illustrates a common MS scenario where the generated code produces an \texttt{E0597} error (``borrowed value does not live long enough'').

A naive compiler feedback approach would simply pass the raw message \texttt{local\_vec dropped here while still borrowed} to the LLM. 
This typically causes the LLM to apply a local patch, such as inserting \texttt{.clone()} inappropriately, which may degrade performance or fail structurally. 
In contrast, our Strategy-Driven RAG maps \texttt{E0597} to the Memory-Semantic (MS) category. 
It retrieves the \texttt{fix\_ownership.j2} template ($\mathcal{T}_i$) and injects the formal definition of Rust lifetime invariants ($\mathcal{K}_i$) into the prompt. 
The augmented prompt serves as a semantic anchor that restricts the stochastic search space to the domain of memory-safe Rust invariants. 
This constraint prevents the LLM from relying on unsafe workarounds and steers it toward an idiomatic structural solution (e.g., expanding the scope of \texttt{local\_vec} or utilizing \texttt{Rc<Vec<i32>>}).

\subsection{Phase 3: Error-Stratified Transformation Strategy (ESTS)}
Phase 3 implements the Error-Stratified Transformation Strategy (ESTS), which adapts repair behavior based on the error category identified during Phase 1. ESTS integrates three mechanisms: (1) diagnostic classification into the four categories defined in Section~\ref{subsec:taxonomy}, (2) per-category temperature scheduling that maps each error stratum to an appropriate level of sampling diversity, and (3) a Local-Loop stagnation escape that resets the repair trajectory when same-category errors persist. 
The error category also drives template selection in Phase 2, so the ESTS classifier serves as the central signal for both knowledge retrieval and sampling control. 
The per-category temperature schedule differs from the static approaches (e.g., $T=0.0$ for greedy decoding or $T=1.0$ for default sampling) used in traditional code generation, which cannot adapt to the semantic characteristics of different compiler signals.
\begin{table}[t]
\centering
\caption{Entropy-Rigidity Mapping in ESTS}
\label{tab:ests_mapping}
\begin{tabular}{@{}llll@{}}
\toprule
\textbf{Error Category} & \textbf{Constraint Type} & \textbf{Target Strategy} & \textbf{Temperature Level} \\ \midrule
SL (Syntactic-Linking)  & Rigid / Deterministic    & Exploitation (Precision) & Low Entropy            \\
MS (Memory-Semantic)    & Semantic / Logical       & Balanced (Multi-path Reasoning) & Moderate Entropy       \\
LB (Logic-Behavioral)   & Loose / Behavioral       & Exploration (Mutation)   & High Entropy           \\
AF (Ambiguous-Fallback) & Unseen / Fallback        & Generalized Exploration  & Baseline Entropy       \\ \bottomrule
\end{tabular}
\end{table}

Table~\ref{tab:ests_mapping} summarizes the Entropy-Rigidity mapping that underpins the temperature schedule. The effectiveness of ESTS stems from the ordering of temperature tiers (low, moderate, and high entropy) rather than the exact parameter values. 
The default parameter values ($\theta_{SL}=0.1$, $\theta_{MS}=0.5$, $\theta_{LB}=1.2$) are selected based on the directional entropy-rigidity mapping and validated through a one-at-a-time sensitivity analysis (Section~\ref{subsec:temp_sensitivity}). This analysis provides diagnostic support that the framework appears reasonably stable under moderate perturbations within each entropy tier.

\begin{itemize}
    \item \textbf{Syntax Precision (Low Entropy)}: 
    Rigid, deterministic SL errors represent a point-to-point repair scenario where the search space is minimal. 
    The goal is to enforce exploitation and minimize invalid paths. We avoid a purely greedy approach ($\theta=0.0$) because zero-entropy sampling frequently traps the LLM in repetition loops. A minimal stochasticity ($\theta_{SL}=0.1$) introduces just enough low-entropy noise to break these loops while preventing deviations from the rigid syntactic constraints.
    \item \textbf{Semantic Reasoning (Moderate Entropy)}: 
    For semantic ownership and lifetime violations (MS), the system adopts a balanced temperature. Resolving Rust ownership trees requires a balance between exploitation (retaining existing logic) and exploration (restructuring the tree). 
    Grounded in recent findings on Self-Consistency in LLMs~\cite{self_consistency}, a moderate entropy provides the model with a multi-path reasoning space to explore diverse resolution strategies without fabricating nonexistent library functions.
    \item \textbf{Algorithmic Exploration (High Entropy)}: 
    Behavioral LB errors indicate that the algorithmic logic has diverged from the source program. The current state is trapped in an incorrect attractor (local optimum). 
    In the framework of Search-Based Software Engineering (SBSE) theory~\cite{sbse}, this high temperature functions analogously to a genetic mutation operator that injects \textit{stochastic perturbation} into the high-dimensional solution space. 
    Such high-entropy disruption is necessary to discard localized patches and escape the current attractor.
    Zhu et al.~\cite{temp_sensitivity} show that dynamically raised temperature for challenging code tokens expands the diversity of generated candidates, which confirms the benefit of elevated entropy for difficult synthesis steps. 
    Based on this finding, we set a high-entropy threshold ($\theta_{LB}=1.2$) to trigger significant algorithmic shifts without loss of structural coherence.
\end{itemize}

\begin{figure}[htbp]
  \centering
  \includegraphics[width=0.92\columnwidth]{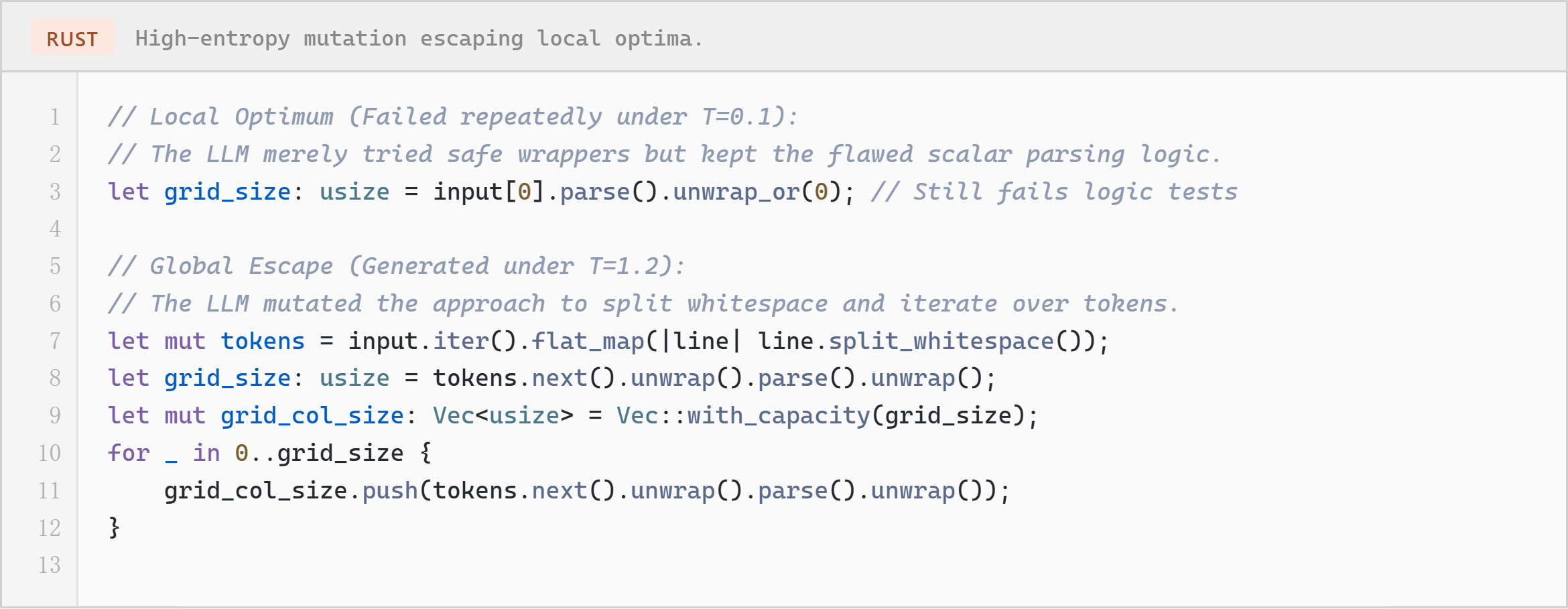}
  \Description{Two Rust code listings shown before and after a high-temperature mutation, where the parsing logic is refactored to escape a local optimum
   and produce a correct solution.}
  \caption{High-entropy mutation escaping local optima by refactoring the parsing logic.}
  \label{lst:mutation_example}
\end{figure}

For instance, in our empirical evaluation (transforming a dynamic programming algorithm), the initial transformation compiled successfully but panicked at runtime (\texttt{ParseIntError}) as shown previously in Fig.~\ref{lst:runtime_panic}. 
A static, low-temperature prompt repeatedly generated minor, ineffective edits that failed the exact same test case. 
A dynamic shift to high entropy alongside a reflection prompt enabled the model to escape this local optimum. 
As demonstrated in Fig.~\ref{lst:mutation_example}, the high entropy induced a complete refactoring of the input parsing mechanism and variable tracking logic. 
This refactored solution ultimately passed the fuzzing suite.

The optimizer also incorporates a \textbf{Local-Loop Stagnation Escape} mechanism to prevent infinite repair loops. 
When the LLM repeatedly produces code that triggers the same error category for more than $N_L$ consecutive iterations (i.e., $\kappa_{local} > N_L$, default $N_L = 3$), the optimizer interprets this as a greedy deadlock. The local repair trajectory has converged to a non-productive attractor. 
The system then fires a RESET that discards the failing Rust draft and re-derives a fresh candidate from the C source at baseline temperature $T = 1.0$. 
The choice $N_L = 3$ is supported by an empirical sweet-spot analysis and a threshold sensitivity study (Section~\ref{sec:evaluation}). Both indicate that the conditional probability of subsequent success drops sharply once the same error category repeats four or more times. 
The selected temperature and the augmented prompt are passed to the LLM, which generates a new candidate program. 
This candidate re-enters Phase 1 for validation, which closes the iterative repair loop.

\subsection{Algorithm Integration: The Closed-Loop Repair Cycle}
Algorithm~\ref{alg:adatrans_main} presents the global execution flow of the AdaTrans framework, which integrates the Multi-Stage Validation (Phase 1), the Strategy-Driven RAG (Phase 2), and the ESTS module (Phase 3).

\begin{algorithm}[tbp]
\caption{AdaTrans Iterative Repair with ESTS}
\label{alg:adatrans_main}
\begin{algorithmic}[1]
\Require Source C code $S_C$; initial Rust program $S_R^{(0)}$; max iterations $M$
\Ensure Optimized Rust code $S_R^*$ or \textsc{Failure}
\State $t \gets 0$; $\kappa_{local} \gets 0$
\While{$t < M$}
    \Statex \hspace{\algorithmicindent}\textit{// Phase 1: Multi-Stage Validation}
    \State $\text{Status}, E^{(t)} \gets \text{Validate}(S_R^{(t)})$
    \If{$\text{Status} = \textsc{Success}$}
        \State \Return $S_R^{(t)}$
    \EndIf
    \State $\text{Category} \gets \text{Classify}(E^{(t)}) \in \{\text{SL}, \text{MS}, \text{LB}, \text{AF}\}$
    \Statex
    \Statex \hspace{\algorithmicindent}\textit{// Update Local-Loop counter}
    \If{$t > 0$ \textbf{and} $\text{Category} = \text{Category}^{(t-1)}$}
        \State $\kappa_{local} \gets \kappa_{local} + 1$
    \Else
        \State $\kappa_{local} \gets 1$
    \EndIf
    \Statex
    \Statex \hspace{\algorithmicindent}\textit{// Local-Loop stagnation escape}
    \If{$\kappa_{local} > N_L$}
        \State $T \gets 1.0$; $\text{Prompt} \gets \mathcal{T}_{RESET}$
        \State $\kappa_{local} \gets 0$
    \Else
        \Statex \hspace{\algorithmicindent}\hspace{\algorithmicindent}\textit{// Phase 2: Strategy-Driven RAG}
        \State $\mathcal{T}_i, \mathcal{K}_i \gets f_{retrieve}(E^{(t)})$
        \State $\text{Prompt} \gets \text{ConstructPrompt}(S_R^{(t)}, \mathcal{T}_i, \mathcal{K}_i)$
        \Statex \hspace{\algorithmicindent}\hspace{\algorithmicindent}\textit{// Phase 3: ESTS temperature scheduling}
        \State $T \gets \mathcal{F}_{ESTS}(\text{Category})$
    \EndIf
    \Statex
    \State $S_R^{(t+1)} \gets \text{LLM.generate}(\text{Prompt},\; T)$
    \State $t \gets t + 1$
\EndWhile
\State \Return \textsc{Failure}
\end{algorithmic}
\end{algorithm}

This algorithm serves as the central orchestrator. 
It highlights how the diagnostic signal $E^{(t)}$ from the deterministic oracle drives both the retrieval function $f_{retrieve}$ and the entropy mapping $\mathcal{F}_{ESTS}$. 
The stagnation escape block (the conditional check on $\kappa_{local}$) implements the Local-Loop fallback mechanism, and ensures that the stochastic search does not fall into an infinite deadlock when the same error category repeats.

%% file: sections/experimental-setup.tex
\section{Experimental Setup}\label{sec:setup}

\subsection{Dataset Construction}\label{subsec:dataset}

The evaluation of C-to-Rust transformation tools in our file-level setting requires a dataset that satisfies three criteria. 
First, the problems should reduce contamination risk with respect to the evaluated LLM snapshot. 
Second, the difficulty spectrum should be sufficient to stress-test both syntactic and semantic transformation capabilities. Third, each problem should have a machine-verifiable execution oracle for automated functional equivalence checking.
Existing resources~\cite{codex,evotrans} often rely on code snippets of uncertain provenance or problems that may overlap with LLM training data. 
Such overlap makes it difficult to distinguish genuine transformation capability from memorized patterns.

We therefore constructed FuzzForLeetcode,\footnote{FuzzForLeetcode: \url{https://github.com/SlainTroyard/FuzzForLeetcode_dev}} a dataset of 104 C programming problems sourced from LeetCode Weekly Contests 413 through 438 (26 contests, 4 problems per contest). 
We selected contests released after September 2024, which postdates the knowledge cutoff (October 2023) of the LLM snapshot used in our experiments (\texttt{gpt-4o-mini-2024-07-18}). 
This temporal separation mitigates, but does not eliminate, the risk that the evaluated model has memorized similar solution patterns during pretraining. 
LeetCode Weekly Contests span a graduated difficulty range from Easy to Hard and cover diverse algorithmic paradigms including dynamic programming, graph traversal, combinatorial optimization, and string processing. 
While these tasks do not capture the full complexity of project-level migration, they provide a controlled benchmark for studying file-level transformation of self-contained programs with executable oracles.

The dataset construction followed a three-stage pipeline:

\textbf{Stage 1: Problem acquisition.}
For each of the 104 problems, we downloaded the problem specification, including input/output format, constraints, and examples. 
Where official C solutions were available from the LeetCode submission archive, we adopted them directly. 
For problems lacking C submissions, a researcher manually implemented a correct C solution following the standard I/O interface (\texttt{scanf}/\texttt{printf}) and verified it by submission to the LeetCode online judge. 
This manual intervention was necessary for approximately 30\% of the problems.

\begin{figure}[htbp]
  \centering
  \includegraphics[width=0.92\columnwidth]{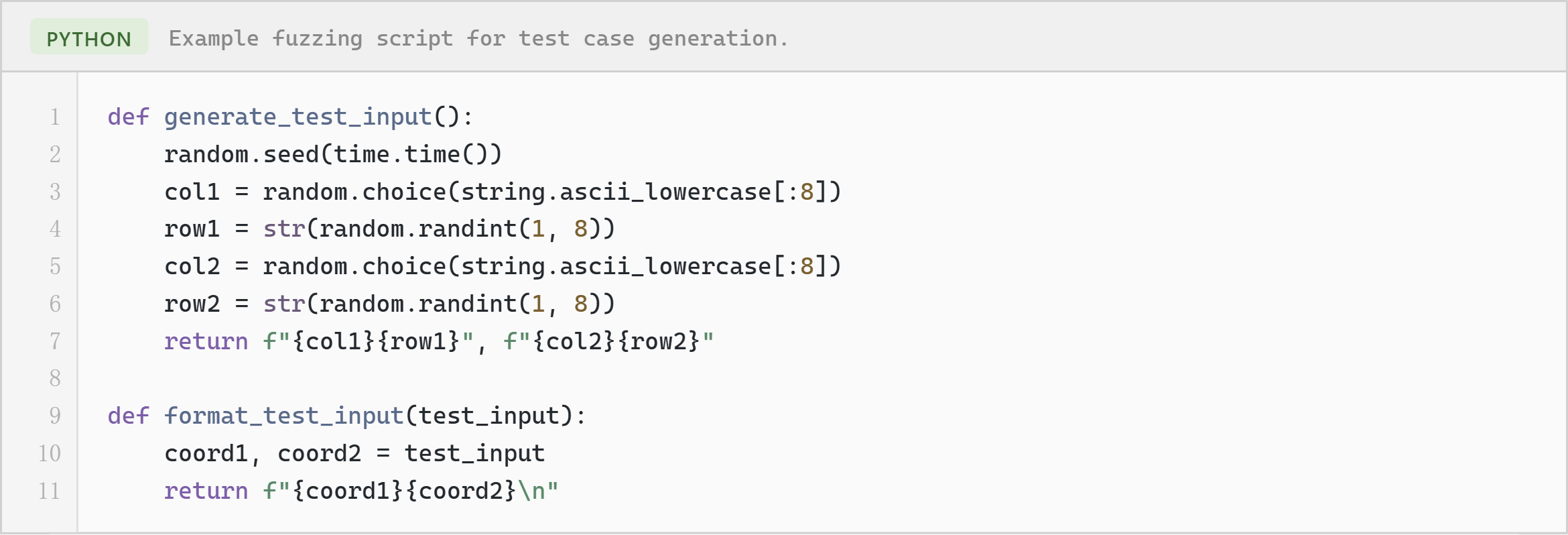}
  \Description{A Python fuzzing script that encodes problem-specific input constraints and randomly generates test cases, then compiles and runs the C
  solution to record reference outputs.}
  \caption{Example fuzzing script for test case generation.}
  \label{lst:fuzz_manual_example}
\end{figure}

\textbf{Stage 2: Fuzzing infrastructure.}
For each problem, we developed a fuzzing script that encodes the input constraints of each problem (e.g., array length bounds, value ranges, string character sets) and programmatically generates random test cases. 
Fig.~\ref{lst:fuzz_manual_example} illustrates the structure of such a script. Each script is manually crafted for the specific input structure of its problem (e.g., arrays, graphs, trees, strings) and generates between 10 and 200 test cases depending on problem complexity, with 100 cases for the majority of problems. 
The generated cases cover both typical inputs and boundary conditions (e.g., minimum/maximum array lengths, extreme values). 
The script then compiles the C solution, feeds each test case via stdin, and records the stdout output.

\textbf{Stage 3: Oracle construction.}
We executed each fuzzing script against its corresponding C solution to produce a deterministic I/O oracle: a set of input--output pairs $(i, o)$ where $o = \mathrm{Out}(S_C, i)$. 
This oracle serves as the ground truth for evaluating functional equivalence of transformed Rust code. 
Any Rust program that produces matching outputs for all test cases (after normalization) is considered functionally equivalent to the original C program. 
This evaluation protocol follows the principle of differential testing~\cite{difftesting}. The original C program and the transformed Rust program are executed on the same randomly generated inputs, and output discrepancies indicate transformation errors. 
The approach is analogous to how Csmith~\cite{csmith_a}~\cite{csmith_b} reveals compiler bugs by comparing the outputs of differently compiled versions of the same program.
In our setting, the C solution plays the role of the trusted reference against which the Rust output is compared.

The resulting dataset comprises 104 problems across 26 weekly contests (4 problems per contest) and spans three official LeetCode difficulty levels: 22 Easy, 48 Medium, and 34 Hard. 
The problems collectively cover a wide algorithmic spectrum under a consistent evaluation protocol. 

\subsection{Baselines and Comparison Models}\label{subsec:baselines}

We compare AdaTrans against the following baseline methods and ablation variants:

\begin{itemize}
    \item \textbf{Zero-shot LLM}: A baseline that transforms C code to Rust using \texttt{gpt-4o-mini} without any repair loop, iterative feedback, or retrieval-augmented context. We sample 200 independent transformations per problem and report pass@$k$ at $k \in \{1, 10, 20, 100\}$. 
    The setting $k = 20$ matches the AdaTrans iteration budget $M = 20$. This enables a budget-matched comparison between independent sampling and iterative repair.

    \item \textbf{AdaTrans (full framework)}: Our complete system integrating Strategy-Driven RAG retrieval, ESTS (error categorization, per-category temperature scheduling, and Local-Loop stagnation detection), and compiler-feedback repair.

    \item \textbf{AdaTrans w/o RAG}: An ablation variant that retains ESTS (error categorization, per-category temperature scheduling, and Local-Loop stagnation reset) but replaces every specialized repair template with a generic prompt that contains the previous Rust draft and the compiler error only. 
    The RVFG doc-tip injection is also disabled. This variant is designed to assess the contribution of the Strategy-Driven RAG module.

    \item \textbf{AdaTrans w/o ESTS}: An ablation variant that retains RAG (specialized templates and doc-tip injection) but removes the entire ESTS module, namely error categorization, per-category temperature scheduling, and Local-Loop stagnation reset. 
    All non-initial repair iterations are routed through a single uniform RAG-augmented template at fixed temperature $T = 1.0$. This variant is designed to assess the contribution of the ESTS module.

    \item \textbf{AdaTrans w/o ESTS \& RAG}: An ablation variant that removes both the ESTS and RAG modules and retains only the iterative repair loop with compiler feedback. 
    All non-initial repair iterations use a generic prompt at fixed temperature $T = 1.0$, without error categorization, temperature scheduling, stagnation reset, specialized templates, or doc-tip injection. 
    This variant also stops iterating as soon as the code compiles, without executing test cases during the repair loop. Only the final output is validated against the fuzz oracle. 
    This reduced variant retains only the basic iterative compilation repair loop and serves as a baseline for the two core innovations of AdaTrans.
\end{itemize}

We also evaluate three existing C-to-Rust transformation tools on the same 104-problem dataset:
\begin{itemize}
    \item \textbf{EvoC2Rust}~\cite{evotrans}: An LLM-based framework that translates C projects through a skeleton-guided strategy. It first generates a compilable Rust skeleton of type-checked function stubs, then incrementally translates each function into its stub, and finally repairs compilation errors by combining the LLM with static analysis. 
    We adapted the authors' provided implementation for file-level evaluation by developing batch orchestration scripts and configuring the LLM client for our evaluation infrastructure. 
    The original implementation was designed for multi-file C projects with build systems and header dependencies. Our adaptation applies its pipeline to standalone single-file LeetCode solutions.
    \item \textbf{Tymcrat}~\cite{tymcrat}: An LLM-based type-migration tool that translates each C function into a Rust function with proper Rust type signatures and iteratively repairs the resulting type errors using Rust compiler feedback.
    We adapted the authors' provided implementation for our dataset, which required resolving numerous crash-inducing edge cases in the transformation engine (e.g., signature parsing panics, unhandled type transformation failures, and keyword collisions with Rust identifiers).
    \item \textbf{PtrTrans}~\cite{ptrtrans}: A project-level C-to-Rust transformation framework that constructs a C-Rust Pointer Knowledge Graph to encode pointer ownership, mutability, nullability, and lifetime information. 
    The enriched pointer semantics are injected into LLM prompts to guide the generation of ownership-compliant Rust code. 
    We adapted the authors' provided implementation for our file-level evaluation by packaging each standalone C solution as a single-file project with the required build metadata.
\end{itemize}
All three tools primarily target project-level transformation. 
We nevertheless evaluate them under the same file-level compilation and I/O consistency protocol used for AdaTrans so that all methods are compared on a common controlled benchmark. 
This adaptation may disadvantage tools whose intended setting is multi-file migration, and we therefore interpret the comparison as evidence within a shared file-level evaluation setup rather than as a definitive judgment of each tool in its original deployment scenario. 
Additionally, we report unsafe code statistics for c2rust~\cite{c2rust}, a traditional AST-based transpiler, on the same dataset.

\textbf{LLM configuration.}
All LLM-based methods (AdaTrans, its ablation variants, and the zero-shot baseline) use the same dated model snapshot, \texttt{gpt-4o-mini-2024-07-18}~\cite{openai2024gpt4ominidocs} (knowledge cutoff: October 2023).
We intentionally fix a single backbone throughout the study to isolate the effect of the AdaTrans algorithmic components from differences in underlying model capability. 
This design also makes full-dataset repeated runs, ablations, and sensitivity analyses feasible within a reproducible evaluation budget. 
We therefore do not claim that the reported gains automatically transfer to all LLM backbones. 
The base sampling parameters are $\mathrm{top\_p} = 1.0$ and $\mathrm{max\_tokens} = 4096$. The base temperature is $T = 1.0$. 
For AdaTrans and the \emph{w/o RAG} ablation, the ESTS module overrides this value with per-category temperatures ($\theta_{SL} = 0.1$, $\theta_{MS} = 0.5$, $\theta_{LB} = 1.2$) as described in Section~\ref{sec:method}. 
The \emph{w/o ESTS} ablation and the zero-shot baseline use the base temperature $T = 1.0$ for all iterations. 
Full prompt templates and the error knowledge base are available in the replication package~\cite{adatrans_replication}.

\subsection{Evaluation Metrics}\label{subsec:metrics}

We adopt the following metrics to evaluate the quality and safety of the transformed Rust code:

\begin{itemize}
    \item \textbf{Compilation Pass Rate}: The percentage of the 104 problems for which the final output compiles successfully without errors. 
    For iterative methods (AdaTrans and its ablation variants), this is the compilation status of the program returned by the optimizer at termination. 
    A higher compilation pass rate indicates that the tool produces syntactically and type-correct Rust code.

    \item \textbf{Functional Equivalence (I/O Consistency) Rate}: The percentage of problems where a compiled Rust program produces output that exactly matches the expected output across all test cases in the evaluation oracle $\mathcal{I}$. 
    Output comparison applies whitespace normalization (stripping leading/trailing whitespace and collapsing internal whitespace) to account for benign formatting differences. 
    All problems in our dataset produce exact integer or string outputs, so floating-point tolerance and order-independent matching are not required. This metric serves as the primary measure of transformation correctness.

    \item \textbf{Unsafe Block Usage}: We measure safety along two dimensions: (i)~the \emph{unsafe file rate}, defined as the percentage of transformed files that contain at least one \texttt{unsafe} block, and (ii)~the \emph{unsafe LOC rate}, defined as the percentage of total generated lines of code that reside within \texttt{unsafe} blocks, aggregated across all output files per method. 
    These metrics quantify whether a transformation tool uses the Rust ownership system or falls back to unsafe abstractions.

    \item \textbf{pass@$k$}~\cite{codex}: For the zero-shot LLM baseline, we report the unbiased estimator of the probability that at least one of $k$ randomly drawn samples passes all test cases. 
    Given $n = 200$ total samples per problem with $c$ correct samples, pass@$k$ is computed as $\mathrm{pass}@k = 1 - \binom{n-c}{k} / \binom{n}{k}$. 
    This estimator follows the standard formulation from Chen et al.~\cite{codex}.

\end{itemize}

%% file: sections/experimental-evaluation.tex
\section{Experimental Evaluation}\label{sec:evaluation}

This section presents the empirical evaluation of AdaTrans to answer the following Research Questions (RQs):

\begin{itemize}
    \item \textbf{RQ1 (Effectiveness): How does AdaTrans perform in terms of compilation pass rate and functional equivalence compared to baseline transformation methods?} 
    To answer this, we evaluate AdaTrans against existing C-to-Rust tools and a zero-shot LLM baseline on a 104-problem dataset. We then report both compilation rates and fuzz-validated solve rates, supplementing these metrics with a difficulty-stratified evaluation and failure mode analysis.
    
    \item \textbf{RQ2 (Safety): To what extent does AdaTrans eliminate the reliance on \texttt{unsafe} blocks to ensure memory safety in the transformed Rust code?} 
    We assess this by measuring the proportion of generated files that contain \texttt{unsafe} blocks and the proportion of total lines of code (LOC) that reside within them, comparing the outputs of our framework against a traditional AST-level transpiler and the zero-shot baseline.
    
    \item \textbf{RQ3 (Ablation): What is the individual contribution of the Strategy-Driven RAG and the ESTS components to the overall performance of the framework?} 
    We investigate this through a macro-level ablation study by systematically disabling these components. Furthermore, we conduct micro-level sensitivity analyses on key experimental configurations, specifically evaluating different per-category temperature settings and local-loop stagnation reset thresholds.
    
    \item \textbf{RQ4 (Convergence): How efficiently does the iterative repair cycle of AdaTrans converge to a valid solution, and how does the error distribution evolve during this process?} 
    We track how the prevalence of the four semantic error categories evolves across successive repair iterations at the dataset level. We also present a detailed case study that traces the repair trajectory of a complex algorithm to illustrate how the adaptive strategies drive convergence.
\end{itemize}

\subsection{RQ1: Effectiveness Analysis}\label{subsec:rq1}

Table~\ref{tab:rq1_overall} presents the overall performance comparison between AdaTrans and the baseline methods on the 104-problem dataset.

\begin{table}[tbp]
\centering
\caption{Overall performance comparison on the 104-problem dataset. AdaTrans and its ablation variants report means across three independent runs. pass@$k$ for $k \in \{1, 10, 20, 100\}$ for the zero-shot LLM are estimated over 200 independent samples.}
\label{tab:rq1_overall}
\begin{tabular}{@{}lcc@{}}
\toprule
\textbf{Method} & \textbf{Compile Rate} & \textbf{Solve Rate} \\ \midrule
c2rust~\cite{c2rust}                   & 96.15\%  & 96.15\% \\
EvoC2Rust~\cite{evotrans}              & 25.00\%  & 1.92\%  \\
Tymcrat~\cite{tymcrat}                 & 81.73\%  & 21.15\% \\
PtrTrans~\cite{ptrtrans}               & 94.23\%  & 10.58\% \\
Zero-shot LLM (pass@1)                 & 57.00\%  & 25.18\% \\
Zero-shot LLM (pass@10)                & 90.78\%  & 56.53\% \\
Zero-shot LLM (pass@20)                & 94.67\%  & 61.48\% \\
Zero-shot LLM (pass@100)               & 98.98\%  & 70.58\% \\
AdaTrans w/o ESTS \& RAG                & 96.79\%  & 46.79\% \\
AdaTrans w/o ESTS                       & 97.76\%  & 69.23\% \\
AdaTrans w/o RAG                        & 98.72\%  & 71.15\% \\ \midrule
\textbf{AdaTrans (Full)}               & \textbf{95.51\%} & \textbf{81.09\%} \\ \bottomrule
\end{tabular}
\end{table}

Across three independent runs, AdaTrans achieves a mean compile rate of 95.51\% $\pm$ 1.11\% and a mean solve rate of 81.09\% $\pm$ 3.09\% (standard deviation), with individual run solve rates of 84.62\%, 79.81\%, and 78.85\%. 
All solve rates in this paper are validated against a fuzz oracle comprising 10--200 test cases per problem, generated independently of the lightweight examples used for iteration feedback.
The Zero-shot LLM pass@k metrics represent distinct usage scenarios. 
The pass@1 value reflects single-shot performance (one API call, no retries) and approximates default practitioner usage. 
The pass@10 and pass@20 values represent moderate resampling effort (10 and 20 independent attempts), and pass@20 matches the AdaTrans iteration budget $M=20$. 
The pass@100 value provides a high-budget reference point for brute-force independent sampling under the same backbone.
AdaTrans achieves its result through a single iterative repair trajectory, a process distinct from brute-force independent sampling.

\textbf{Compilation Pass Rate.}
Among existing tools, c2rust achieves a compile rate of 96.15\% (100/104) when evaluated on the nightly toolchain that its generated \texttt{extern\_types} declarations require, followed by PtrTrans at 94.23\% (98/104), Tymcrat at 81.73\% (85/104), and EvoC2Rust at 25.00\% (26/104). 
The four remaining c2rust compilation failures occur because the C source uses GCC-extension nested function definitions (three problems) or variable-length array initializers (one problem), constructs for which c2rust omits the affected function bodies entirely. 
The high compilation success of PtrTrans stems from its SVF-based pointer analysis, which injects precise ownership, mutability, and lifetime annotations into the LLM prompt. 
However, as discussed below, this compilation advantage does not extend to functional correctness. 
The low compilation rate of EvoC2Rust stems from an architectural mismatch: it is designed for project-level C-to-Rust transformation involving multi-file codebases with headers, build systems, and inter-module dependencies.
Its pipeline (extract metadata, generate Rust skeleton, incrementally transform functions, then repair) introduces failure points when applied to standalone single-file programs. 
Specifically, EvoC2Rust generates code that relies on undefined convenience macros such as \texttt{c\_qsort!}, \texttt{c\_malloc!}, \texttt{c\_sizeof!}, and \texttt{c\_for!}. 
These macros are part of its transformation utility library but are not consistently included in the generated output. 
Its type mapping is also imprecise for isolated functions. For example, it produces \texttt{i64} where \texttt{i32} is expected, which causes systematic \texttt{mismatched types} errors. 
Its repair chain (bracket fix $\rightarrow$ rule fix $\rightarrow$ LLM repair) was designed for large-project compilation errors rather than the systematic type mismatches and missing definitions that arise from single-function transformation. 
We note that the reported results (25.00\% compile, 1.92\% solve) are after adapting the EvoC2Rust infrastructure for our evaluation pipeline. 
The original implementation produced even lower success rates due to hardcoded API configurations and missing batch evaluation support.

AdaTrans achieves a mean compile rate of 95.51\%, comparable to c2rust (96.15\%) and surpassing Tymcrat by 13.78 percentage points and EvoC2Rust by 70.51 percentage points. 
The Zero-shot LLM pass@100 compilation estimate reaches 98.98\%, but this metric reflects the near-certainty of finding at least one compiling sample among 100 independent draws rather than a per-attempt success rate. 
AdaTrans achieves this compilation rate through a single iterative repair trajectory that systematically addresses compiler errors across multiple repair attempts. 
This approach avoids the cost of generating and evaluating hundreds of candidate programs. 
The Strategy-Driven RAG module further contributes by steering repairs toward functionally correct solutions (Section~\ref{subsec:rq3}).

\textbf{Functional Equivalence (Solve Rate).}
Under our file-level evaluation protocol, AdaTrans achieves a mean solve rate of 81.09\% and surpasses Tymcrat (21.15\%) by 59.94 percentage points, PtrTrans (10.58\%) by 70.51 percentage points, and EvoC2Rust (1.92\%) by 79.17 percentage points. 
c2rust achieves the highest solve rate among all tools at 96.15\% (100/104) because its deterministic AST transpilation preserves C semantics faithfully whenever the source compiles. 
Every c2rust file that compiles also passes all fuzz tests, with a compile-to-solve gap of zero. 
However, this functional fidelity comes at the cost of memory safety. As shown in Table~\ref{tab:rq2_safety}, 100\% of the c2rust output files contain \texttt{unsafe} blocks (97.94\% of generated LOC), effectively forfeiting the safety guarantees that motivate C-to-Rust migration. 
For practitioners who prioritize correctness over safety, c2rust followed by manual \texttt{unsafe} cleanup represents a viable alternative workflow. 
AdaTrans targets the complementary scenario where memory safety is a first-class requirement. 
AdaTrans achieves a mean solve rate of 81.09\% with a mean \texttt{unsafe} file rate of 1.19\%, a result that shows that iterative repair with semantic knowledge retrieval can achieve high functional equivalence rates while producing code that satisfies the Rust ownership guarantees.

The large compile-to-solve gap of Tymcrat (81.73\% compile vs.\ 21.15\% solve) indicates that type-directed mutation alone is insufficient for functional correctness.
While Tymcrat can map C types to Rust equivalents well enough to pass the compiler, it lacks a mechanism to verify and repair algorithmic behavior. 
PtrTrans exhibits an even wider gap (94.23\% compile vs.\ 10.58\% solve), the largest among all tools. 
Although its SVF-based pointer analysis produces code that satisfies Rust ownership constraints, the transformed programs diverge from the original C semantics in most cases. 
Precise pointer ownership inference thus addresses the type-level challenge of C-to-Rust transformation but not the algorithmic fidelity required for functional equivalence. 
The low functional equivalence rate of EvoC2Rust (1.92\%) reflects the engineering challenges of adapting a project-level pipeline to file-level evaluation.

AdaTrans surpasses the Zero-shot LLM pass@1 rate (25.18\%) by 55.91 percentage points. 
When the Zero-shot LLM is allowed ten or twenty attempts (pass@10 = 56.53\%, pass@20 = 61.48\%), with pass@20 matching the AdaTrans iteration budget $M=20$, AdaTrans still leads by 24.56 and 19.61 percentage points respectively. 
The single iterative repair trajectory of AdaTrans also surpasses the Zero-shot LLM pass@100 estimate (70.58\%) by 10.51 percentage points. 
This result suggests that feedback-guided repair can be more effective than repeated independent sampling in this setting. 
AdaTrans consumes an average of \textbf{11{,}545 tokens per problem} across all 104 problems, while the zero-shot baseline requires approximately 31{,}875 tokens per problem for 20 independent samples. 
AdaTrans thus achieves a higher mean solve rate (\textbf{81.09\% vs.\ 61.48\%}) at roughly \textbf{one-third the token cost}.
Overall, AdaTrans processes the 104-problem dataset at an estimated cost of \$0.44 USD (629K prompt tokens + 571K completion tokens at \texttt{gpt-4o-mini} pricing). This corresponds to \$0.004 per problem and 5.47 iterations per problem on average.

The gap between compile rate and solve rate reveals the inherent difficulty of achieving functional equivalence in C-to-Rust transformation. 
While syntactic and ownership violations can be systematically diagnosed and repaired through compiler feedback, logic-behavioral errors, where the code compiles but produces incorrect results, require deeper semantic understanding that AdaTrans addresses through its ESTS error categorization and adaptive repair strategy.

\textbf{Difficulty-Stratified Analysis.}
The following diagnostic analyses (difficulty stratification, failure analysis, sensitivity studies, and convergence analysis) are based on one complete run. 
Table~\ref{tab:difficulty_breakdown} stratifies solve rates by LeetCode difficulty level for AdaTrans and the zero-shot LLM baseline.

\begin{table}[tbp]
\centering
\caption{Solve rates by problem difficulty. AdaTrans results are from one complete run. The pass@$k$ values are estimated from 200 independent samples. $n$ denotes the number of problems per difficulty level.}
\label{tab:difficulty_breakdown}
\begin{tabular}{@{}lcccc@{}}
\toprule
\textbf{Difficulty} & $n$ & \textbf{AdaTrans} & \textbf{pass@20} & \textbf{pass@100} \\ \midrule
Easy   & 22  & 95.45\% & 84.48\% & 94.55\% \\
Medium & 48  & 75.00\% & 57.72\% & 64.59\% \\
Hard   & 34  & 91.18\% & 51.92\% & 63.52\% \\ \midrule
Overall & 104 & 84.62\% & 61.48\% & 70.58\% \\
\bottomrule
\end{tabular}
\end{table}

AdaTrans outperforms pass@100 across all three difficulty levels, with the largest advantage on Hard problems (91.18\% vs.\ 63.52\%, a 27.66 percentage-point gap). On Medium problems, AdaTrans achieves 75.00\% versus 64.59\% for pass@100 (10.41~pp). 
The non-monotonic pattern (Hard problems yield a higher solve rate than Medium) indicates that LeetCode difficulty, which reflects algorithmic complexity for human programmers, does not directly predict C-to-Rust transformation difficulty. 
Transformation difficulty depends more on the prevalence of pointer-intensive idioms (e.g., linked lists, manual memory management) than on algorithmic sophistication.

\textbf{Failure Analysis.}
We examine one complete run in detail to illustrate the distribution of failure modes. Of the 104 problems, 16 fail validation. 
These fall into two categories. Seven problems never pass even the lightweight example tests during iteration, and nine problems pass examples but fail when validated against the full fuzz oracle.

The 7 unsolved problems exhibit two distinct failure modes by final compilation status. Four exhaust the 20-iteration budget with a final compilation error, while three compile successfully but produce incorrect output. 
The ESTS diagnostic classification labels each problem by the error category that drove the final repair attempt, which reflects the diagnostic signal of the preceding iteration rather than the final compilation outcome. 
By this classification, logic-behavioral (LB) errors dominate with 6 problems, followed by memory-semantic (MS, 1). Of the 6 LB-classified problems, 4 regressed to compilation errors during their final repair attempt. 
The high-temperature exploration triggered by the LB classification ($\theta_{LB} = 1.2$) occasionally destabilizes previously working syntax while attempting algorithmic restructuring. 
The prevalence of LB classifications indicates that algorithmic reconstruction, not syntax or ownership repair, is the primary bottleneck for the remaining unsolved problems. All 4 compilation failures achieved successful compilation at earlier iterations (compiling in 1--16 of 20 iterations) but regressed after high-temperature exploration or stagnation resets redirected the search.

The 9 false-positive problems pass the 2--3 example test cases used for iteration feedback but fail on the full fuzz oracle (10--200 test cases per problem).
These failures indicate that the generated code overfits to the small example set without capturing the full algorithmic specification. 
Increasing the number of iteration feedback test cases beyond the current 2--3 examples would likely reduce the false positive rate, at the cost of increased execution time per iteration. We leave the exploration of this trade-off to future work. 
The false positives concentrate at contest positions~2 and~4 (3 each), with 2 at position~3 and 1 at position~1, where positions~1--4 correspond to increasing difficulty within each weekly contest (Section~\ref{subsec:dataset}).

The difficulty-stratified analysis (Table~\ref{tab:difficulty_breakdown}) shows that Medium problems have the lowest solve rate among the three difficulty levels, and suggests that Medium-difficulty algorithmic patterns pose the greatest challenge for C-to-Rust transformation.

\subsection{RQ2: Safety Analysis}\label{subsec:rq2}

A central promise of the Rust programming language is the guarantee of memory safety without runtime overhead~\cite{rust_study_1}~\cite{rust_study_2}. 
However, this guarantee is contingent on the absence of \texttt{unsafe} blocks, which bypass the borrow checker and reintroduce the memory vulnerabilities that Rust aims to eliminate~\cite{unsafe_rust}. 
We evaluate the safety of AdaTrans output by measuring both the proportion of transformed files containing \texttt{unsafe} blocks and the proportion of generated LOC residing within \texttt{unsafe} blocks.

Table~\ref{tab:rq2_safety} presents the safety metrics for AdaTrans, the zero-shot LLM baseline, and c2rust.

\begin{table}[tbp]
\centering
\caption{Safety comparison: \texttt{unsafe} block usage in transformed Rust code. The AdaTrans unsafe file rate is the three-run mean. c2rust statistics are computed over 104 transformed files.}
\label{tab:rq2_safety}
\begin{tabular}{@{}lcc@{}}
\toprule
\textbf{Method} & \textbf{Unsafe File Rate} & \textbf{Unsafe LOC Rate$^\dagger$} \\ \midrule
c2rust~\cite{c2rust}                & 100.00\% & 97.94\% \\
Zero-shot LLM (\texttt{gpt-4o-mini})        & 1.05\%   & ${<}\,0.1$\%    \\
\textbf{AdaTrans (Full)}            & \textbf{1.19\%} & $\mathbf{{<}\,0.1}$\textbf{\%} \\ \bottomrule
\multicolumn{3}{@{}p{0.92\columnwidth}@{}}{\footnotesize $^\dagger$ Percentage of total generated LOC within \texttt{unsafe} blocks, aggregated across all output files per method.}
\end{tabular}
\end{table}

AdaTrans introduces \texttt{unsafe} blocks in only 1.19\% of successfully validated programs (three-run mean, with exactly one \texttt{unsafe} file per run), and less than 0.1\% of total generated LOC resides inside \texttt{unsafe} blocks. 
The recurring \texttt{unsafe} case arises in a problem whose original C solution relies on global mutable state (multiple statically allocated lookup tables), and the LLM elected to mirror this pattern via \texttt{static mut} declarations rather than refactoring to interior-mutability primitives such as \texttt{LazyLock} or \texttt{OnceCell}. 
The zero-shot LLM baseline shows a comparable unsafe file rate (1.05\%) and a similarly negligible unsafe LOC rate. 
By contrast, as an AST-level transpiler, c2rust preserves C semantics by wrapping nearly all transformed code in \texttt{unsafe} blocks (97.94\% of generated LOC). 
This wrapping negates the memory safety guarantees that motivate Rust adoption.
c2rust also fails to produce compilable output for 4 of the 104 problems where the C source contains nested function definitions or variable-length array initializers with no direct Rust equivalent. 
Even a single \texttt{unsafe} block can undermine the compiler's ability to guarantee the absence of use-after-free, double-free, and data-race bugs~\cite{unsafe_rust}.

The near-elimination of \texttt{unsafe} code in AdaTrans is a direct consequence of the Strategy-Driven RAG mechanism. 
When the LLM encounters memory-semantic (MS) violations, the RAG module retrieves ownership-aware repair templates ($\mathcal{T}_{MS}$) that guide the model toward idiomatic safe Rust patterns, such as ownership scope restructuring, reference counting (\texttt{Rc}, \texttt{Arc}), or the Rust standard collection APIs, instead of \texttt{unsafe} escape hatches. 
These results indicate that high transformation success rates and memory safety are largely compatible. By combining iterative repair with semantic knowledge retrieval, the framework achieves a mean solve rate of 81.09\% while keeping \texttt{unsafe} usage two orders of magnitude below c2rust.

\subsection{RQ3: Ablation Study}\label{subsec:rq3}

We conduct an ablation study to quantify the individual contribution of each core component, systematically removing the Strategy-Driven RAG and the ESTS modules from the full AdaTrans framework. Table~\ref{tab:rq3_ablation} presents the ablation results.

\begin{table}[tbp]
\centering
\caption{Ablation study: component-wise contribution analysis on the 104-problem dataset. All values are means across three independent runs ($\pm$ standard deviation).}
\label{tab:rq3_ablation}
\begin{tabular}{@{}lccc@{}}
\toprule
\textbf{Variant} & \textbf{Compile Rate} & \textbf{Solve Rate} & \textbf{$\Delta$ Solve (pp)} \\ \midrule
AdaTrans w/o ESTS \& RAG              & 96.79\% $\pm$ 2.00  & 46.79\% $\pm$ 3.89  & $-$34.30 \\
AdaTrans w/o ESTS                     & 97.76\% $\pm$ 1.11  & 69.23\% $\pm$ 7.51  & $-$11.86 \\
AdaTrans w/o RAG                      & 98.72\% $\pm$ 1.47  & 71.15\% $\pm$ 5.85  & $-$9.94 \\ \midrule
\textbf{AdaTrans (Full)}              & \textbf{95.51\%} $\pm$ \textbf{1.11} & \textbf{81.09\%} $\pm$ \textbf{3.09} & --- \\ \bottomrule
\end{tabular}
\end{table}

A notable pattern in Table~\ref{tab:rq3_ablation} is that the full framework has the lowest mean compile rate (95.51\%) among all variants. 
This reflects the cost of high-entropy LB exploration ($\theta_{LB} = 1.2$): aggressive algorithmic restructuring occasionally destabilizes previously compiling code, but the resulting solve-rate gain (up to 34.30~pp) far outweighs the modest compile-rate reduction.

\textbf{Impact of removing both components.}
The ${w/o ESTS \& RAG}$ variant retains only the iterative repair loop with compiler feedback: all errors use a generic prompt at fixed temperature $T = 1.0$, with no error categorization, no stagnation reset, no specialized templates, and no doc-tip injection. 
This variant also stops iterating as soon as the code compiles, without executing test cases during the repair loop, only the final output is validated against the fuzz oracle. 
Across three runs, it achieves a mean solve rate of 46.79\% ($\pm$ 3.89\%), a 34.30 percentage-point drop from the full framework. 
The large gap suggests that the ESTS and RAG components contribute substantially to the full framework's effectiveness beyond what the iterative repair loop alone provides. 
The 46.79\% solve rate still substantially underperforms the zero-shot LLM pass@20 estimate (61.48\%). 
This result indicates that iterative compilation repair without test-case feedback and adaptive strategies is less effective than independent sampling with equivalent budget.

\textbf{Impact of Strategy-Driven RAG.}
Removing the RAG module (${w/o RAG}$) replaces every specialized repair template with a generic prompt that exposes only the previous Rust draft and the compiler error, without doc-tip knowledge injection. 
The mean compile rate remains high at 98.72\% ($\pm$ 1.47\%), which indicates that the LLM can produce syntactically valid Rust without targeted templates. 
The mean solve rate, however, falls from 81.09\% to 71.15\% ($\pm$ 5.85\%), a 9.94 percentage-point reduction. 
This disparity reveals the distinct role of RAG. Raw compiler feedback suffices to drag the program through Rust syntax and ownership checks, but the strategy-driven retrieval of error-category-specific templates is essential for directing the LLM toward functionally correct algorithmic implementations. 
The RAG module provides not only compilation guidance but also idiom-level repair knowledge.

\textbf{Impact of ESTS.}
Removing the ESTS module (${w/o ESTS}$) disables error categorization, the per-category temperature schedule, and the Local-Loop stagnation reset.
All errors are routed through a uniform RAG-augmented template at fixed temperature $T=1.0$. 
The mean compile rate remains high at 97.76\% ($\pm$ 1.11\%), but the mean solve rate drops from 81.09\% to 69.23\% ($\pm$ 7.51\%), an 11.86 percentage-point reduction. 
Two observations emerge from this comparison. 
First, ESTS is the larger contributor to functional correctness in our ablation, with a slightly greater drop than removing RAG. 
Second, the high compile rate of the ablated variants indicates that without categorization, repair iterations gravitate toward conservative, locally-correct edits that compile but fail to recover the intended algorithm.

\textbf{Synergistic Effects.}
The two components are tightly coupled. 
ESTS supplies the error-category signal that selects the appropriate RAG template.
RAG supplies the structured repair guidance that turns the ESTS temperature schedule into productive search. 
Removing either component reduces the mean solve rate by approximately 10--12 percentage points, while removing both causes a 34.30 percentage-point drop, substantially larger than the sum of individual removals would suggest if the components were independent. 
This super-additive degradation suggests that the two modules reinforce each other. Adaptive temperature scheduling in ESTS is most effective when paired with matched RAG templates, and vice versa. 
We note that part of the additional gap may also reflect the absence of test-case feedback during the repair loop in the ${w/o ESTS \& RAG}$ variant, which stops at first successful compilation rather than verifying functional equivalence at each iteration. The synergy estimate is therefore an upper bound.

\subsubsection{Temperature Sensitivity Analysis}\label{subsec:temp_sensitivity}
A natural question is whether ESTS depends critically on the specific temperature values ($\theta_{SL} = 0.1$, $\theta_{MS} = 0.5$, $\theta_{LB} = 1.2$) or on the directional mapping itself, that is, low temperature for rigid syntax fixes, moderate temperature for ownership reasoning, and high temperature for algorithmic exploration.
We investigate this by comparing the default configuration against three alternative settings on the full 104-problem dataset with budget $M = 20$.

\begin{table}[tbp]
\centering
\caption{Temperature sensitivity analysis on the full 104-problem C dataset (one complete run). Each variant overrides all three category temperatures simultaneously. The Local-Loop threshold $N_L = 3$ is held constant.}
\label{tab:temp_sensitivity}
\begin{tabular}{@{}lccc@{}}
\toprule
\textbf{Variant} & $(\theta_{SL}, \theta_{MS}, \theta_{LB})$ & \textbf{Solve} & \textbf{$\Delta$ Solve (pp)} \\ \midrule
\textbf{Default}             & $(0.1, 0.5, 1.2)$ & \textbf{84.62\%} & --- \\
Uniform-Moderate             & $(0.5, 0.5, 0.5)$ & 69.23\%          & $-$15.39 \\
Uniform-High                 & $(1.2, 1.2, 1.2)$ & 74.04\%          & $-$10.58 \\
Reversed                     & $(1.2, 0.5, 0.1)$ & 67.31\%          & $-$17.31 \\
\bottomrule
\end{tabular}
\end{table}

\textbf{Discussion.}
Table~\ref{tab:temp_sensitivity} reveals two findings. First, the differentiated default $(0.1, 0.5, 1.2)$ achieves the highest solve rate among all tested variants. 
The advantage over Uniform-High is 10.58 percentage points, over Uniform-Moderate 15.39 percentage points, and over Reversed 17.31 percentage points. 
The consistent direction supports the entropy-rigidity rationale. 
Syntax errors benefit from low-temperature determinism, while logic errors benefit from high-temperature exploration. 
The Reversed configuration, which assigns high temperature to syntax errors and low temperature to logic errors, suffers the largest drop, which supports the conclusion that the mapping direction matters. 
Second, the ablation study (Section~\ref{subsec:rq3}) shows that the full ESTS module contributes an 11.86 percentage-point gain in mean solve rate over the ${w/o ESTS}$ baseline. 
The temperature sensitivity analysis complements this finding. 
Differentiated per-category temperatures outperform all uniform configurations tested and provide diagnostic support for the directional mapping (low for syntax, high for logic) as a key driver within ESTS.

\subsubsection{Local-Loop Threshold Sensitivity Analysis}\label{subsec:threshold_sensitivity}
ESTS triggers a RESET (fall back to a fresh transformation from the C source) when the same error category repeats more than $N_L$ times in succession. 
We justify the chosen value $N_L = 3$ by evaluating four configurations on the full 104-problem dataset with budget $M = 20$.

\begin{table}[tbp]
\centering
\caption{Local-Loop threshold sensitivity (one complete run). Other ESTS settings (per-category temperatures) are held at their defaults.}
\label{tab:threshold_sensitivity}
\begin{tabular}{@{}lccc@{}}
\toprule
\textbf{Variant} & $N_L$ & \textbf{Compile} & \textbf{Solve} \\ \midrule
Aggressive                       & 1        & 86.54\% & 73.08\% \\
\textbf{Default}                  & \textbf{3} & \textbf{96.15\%} & \textbf{84.62\%} \\
Lazy                              & 5        & 91.35\% & 69.23\% \\
Disabled                          & $\infty$ & 85.58\% & 67.31\% \\
\bottomrule
\end{tabular}
\end{table}

\begin{figure}[htbp]
\centering
\begin{tikzpicture}
\begin{axis}[
    width=0.85\columnwidth,
    height=5cm,
    xlabel={Consecutive same-category errors ($\kappa_{local}$)},
    ylabel={$P(\text{success in next 5 iters})$ (\%)},
    xmin=0.5, xmax=5.5,
    ymin=-2, ymax=60,
    xtick={1,2,3,4,5},
    ytick={0,10,20,30,40,50,60},
    grid=major,
    grid style={dashed, gray!30},
    every node near coord/.append style={font=\scriptsize, anchor=south, yshift=2pt},
    nodes near coords,
]
\addplot[mark=*, mark size=3pt, thick, blue!70] coordinates {
    (1, 51.0) (2, 40.4) (3, 37.5) (4, 9.1) (5, 0.0)
};
\addplot[dashed, red!70, mark=none, forget plot] coordinates {(3.5, -2) (3.5, 60)};
\node[red!70, font=\footnotesize] at (axis cs:3.5, 56) {RESET threshold};
\end{axis}
\end{tikzpicture}
\Description{A line chart plotting conditional success probability against the Local-Loop counter value. The curve decreases from 51.0 percent at
  counter 1 to 37.5 percent at counter 3, then drops sharply to 9.1 percent at 4 and 0.0 percent at 5, with a dashed vertical line marking the RESET
  threshold at 3.5.}
\caption{Sweet-spot analysis of the Local-Loop threshold from AdaTrans repair logs (434 observations across error-category iterations). The conditional success probability decreases monotonically from 51.0\% at $\kappa_{local} = 1$ to 37.5\% at $\kappa_{local} = 3$, then drops sharply to 9.1\% at $\kappa_{local} = 4$ and 0.0\% at $\kappa_{local} = 5$, which supports $N_L = 3$ as the threshold.}
\label{fig:sweet_spot}
\end{figure}
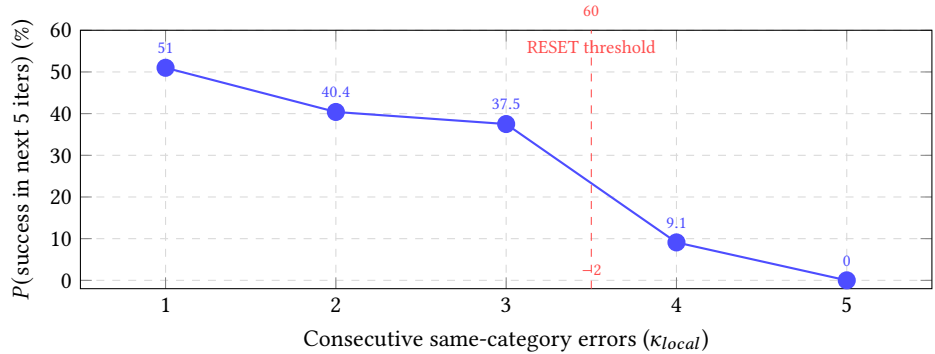

\textbf{Empirical sweet-spot.}
The value $N_L = 3$ is further justified by a post-hoc analysis of the AdaTrans repair logs. 
For each iteration step, we compute the conditional probability that AdaTrans reaches a successful solution within the next five iterations, conditioned on the current value of $\kappa_{local}$. 
Fig.~\ref{fig:sweet_spot} shows the curve. The probability decreases monotonically from 51.0\% at $\kappa_{local} = 1$ to 37.5\% at $\kappa_{local} = 3$, then drops sharply to 9.1\% at $\kappa_{local} = 4$ (11 observations) and 0.0\% at $\kappa_{local} = 5$ (7 observations). 
The steep decline beyond $\kappa_{local} = 3$ indicates that recovery probability drops by a factor of four between the third and fourth consecutive same-category repetition. 
By $\kappa_{local} = 5$, no successful recovery is observed. 
The marginal benefit of allowing a fourth same-category attempt (9.1\%) is substantially outweighed by the opportunity cost of consuming an iteration budget slot on a near-exhausted repair trajectory, which supports $N_L = 3$ as the threshold.

\textbf{Discussion.}
The sensitivity table mirrors the sweet-spot analysis. $N_L = 3$ achieves the highest solve rate (84.62\%). Aggressive RESET ($N_L = 1$) loses 11.54 percentage points by interrupting still-productive repair sequences, while lazy or disabled RESET ($N_L = 5$ at 69.23\%, $N_L = \infty$ at 67.31\%) wastes budget on saturated trajectories. 
The compile-rate trade-off is gentler because abandoning a same-category repair sometimes regresses to a compile-error state on the next iteration. 
The simplified single-trigger design is empirically motivated. An earlier version of AdaTrans included secondary thresholds for ``compilation hell'' ($N_C$) and ``algorithmic deadlock'' ($N_A$). 
However, a log analysis of that multi-trigger prototype shows that the Local-Loop trigger fires before either of the others in 95.9\% of stagnation events (118 of 123 trigger-condition evaluations across the full dataset). 
The two secondary triggers fire only 5 times combined. 
Disabling them entirely yields the configuration adopted in this paper.

\subsection{RQ4: Convergence and Error Repair Analysis}\label{subsec:rq4}

This RQ investigates the convergence behavior of the iterative repair cycle of AdaTrans and the evolution of error categories across successive iterations. 
The analysis in this subsection is based on one complete run. 
We analyze the error distribution through the lens of the four-category taxonomy defined in Section~\ref{subsec:taxonomy}, namely Syntactic-Linking (SL), Memory-Semantic (MS), Logic-Behavioral (LB), and Ambiguous-Fallback (AF).

\subsubsection{Error Distribution in the Knowledge Base}
The RVFG knowledge base encodes 517 distinct Rust compiler error codes. Table~\ref{tab:error_distribution} shows their distribution across the four ESTS categories.

\begin{table}[tbp]
\centering
\caption{Distribution of Rust compiler error codes in the RVFG knowledge base across ESTS categories.}
\label{tab:error_distribution}
\begin{tabular}{@{}lrl@{}}
\toprule
\textbf{Category} & \textbf{Count (\%)} & \textbf{Scope} \\ \midrule
SL (Syntactic-Linking) & 301 (58.2\%) & Type system, module resolution, syntax \\
MS (Memory-Semantic)   & 105 (20.3\%) & Ownership, lifetime, borrow checker \\
LB (Logic-Behavioral)  &   1 \phantom{0}(0.2\%) & Runtime logic (post-compilation) \\
AF (Ambiguous-Fallback) &  9 \phantom{0}(1.7\%) & External macros, crate interactions \\ \midrule
Obsolete               & 101 (19.5\%) & Deprecated or version-specific codes \\
\bottomrule
\end{tabular}
\end{table}

SL errors dominate the knowledge base (58.2\%), which reflects the syntactic rigidity of the Rust type system and module resolution. 
The substantial MS proportion (20.3\%) reflects the unique challenge of mapping manual memory management in C to the Rust ownership model. 
LB errors are rare at the compiler level because logic defects manifest only after successful compilation, during test execution.

\subsubsection{Convergence Behavior}
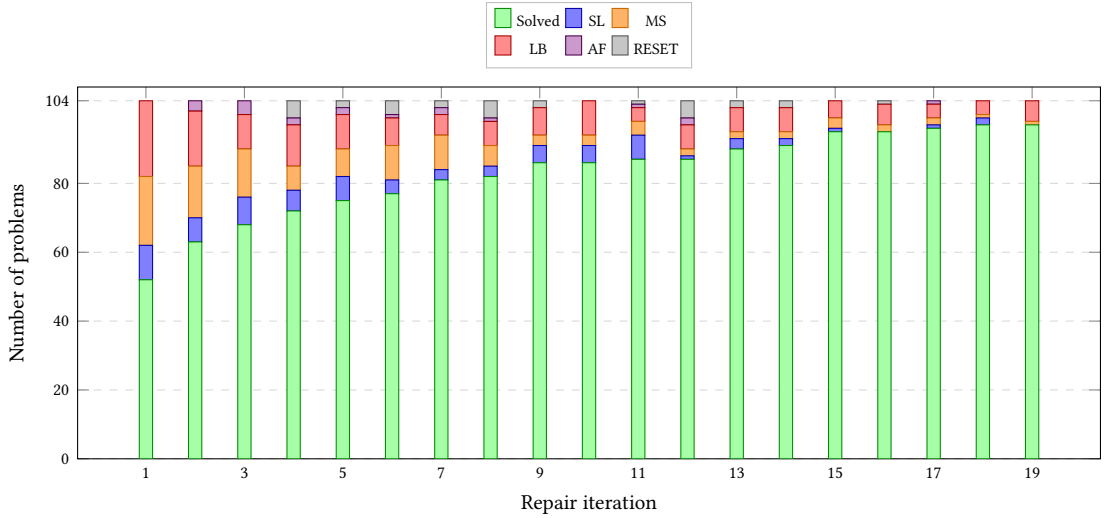
\begin{figure}[htbp]
\centering
\begin{tikzpicture}
\begin{axis}[
    ybar stacked,
    bar width=5pt,
    width=\columnwidth,
    height=6.5cm,
    xlabel={Repair iteration},
    ylabel={Number of problems},
    xlabel style={font=\small},
    ylabel style={font=\small},
    xmin=0.2, xmax=19.8,
    ymin=0, ymax=108,
    xtick={1,3,...,19},
    x tick label style={font=\scriptsize},
    y tick label style={font=\scriptsize},
    ytick={0,20,40,60,80,104},
    ymajorgrids=true,
    grid style={dashed, gray!30},
    legend style={at={(0.5,1.05)}, anchor=south, legend columns=3, font=\scriptsize, draw=gray!50},
    enlarge x limits=0.03,
]
\addplot[fill=green!35, draw=green!55!black] coordinates {
    (1,52) (2,63) (3,68) (4,72) (5,75) (6,77) (7,81) (8,82)
    (9,86) (10,86) (11,87) (12,87) (13,90) (14,91) (15,95)
    (16,95) (17,96) (18,97) (19,97)
};
\addplot[fill=blue!50, draw=blue!70!black] coordinates {
    (1,10) (2,7) (3,8) (4,6) (5,7) (6,4) (7,3) (8,3)
    (9,5) (10,5) (11,7) (12,1) (13,3) (14,2) (15,1)
    (16,0) (17,1) (18,2) (19,0)
};
\addplot[fill=orange!60, draw=orange!80!black] coordinates {
    (1,20) (2,15) (3,14) (4,7) (5,8) (6,10) (7,10) (8,6)
    (9,3) (10,3) (11,4) (12,2) (13,2) (14,2) (15,3)
    (16,2) (17,2) (18,1) (19,1)
};
\addplot[fill=red!45, draw=red!65!black] coordinates {
    (1,22) (2,16) (3,10) (4,12) (5,10) (6,8) (7,6) (8,7)
    (9,8) (10,10) (11,4) (12,7) (13,7) (14,7) (15,5)
    (16,6) (17,4) (18,4) (19,6)
};
\addplot[fill=violet!40, draw=violet!60!black] coordinates {
    (1,0) (2,3) (3,4) (4,2) (5,2) (6,1) (7,2) (8,1)
    (9,0) (10,0) (11,1) (12,2) (13,0) (14,0) (15,0)
    (16,0) (17,1) (18,0) (19,0)
};
\addplot[fill=gray!45, draw=gray!65!black] coordinates {
    (1,0) (2,0) (3,0) (4,5) (5,2) (6,4) (7,2) (8,5)
    (9,2) (10,0) (11,1) (12,5) (13,2) (14,2) (15,0)
    (16,1) (17,0) (18,0) (19,0)
};
\legend{Solved, SL, MS, LB, AF, RESET}
\end{axis}
\end{tikzpicture}
\Description{A stacked bar chart showing the count of problems in each ESTS error category, Solved, SL, MS, LB, AF, and RESET, across repair iterations
   1 to 19. Solved problems accumulate over iterations while the remaining error categories taper off toward the later iterations.}
\caption{Error category distribution across repair iterations (1--19) for all 104 problems. At each iteration, each problem is either Solved (passed feedback tests at a prior iteration) or classified by its ESTS error category for the current repair attempt. RESET indicates Local-Loop stagnation resets ($\kappa_{local} > N_L$). Twenty-five problems (24.0\%) solve on the initial transformation (iteration~0, not shown).}
\label{fig:error_evolution}
\end{figure}

The iterative repair cycle of AdaTrans converges efficiently across the dataset. 
For the 100 problems that achieve compilation success, the majority require only a small number of iterations. 
Actual repair trajectories are non-linear. Errors from different categories can interleave, recur after apparent resolution, or re-emerge following stagnation resets (see Section~\ref{subsec:case_study}). 
Even so, we observe three dominant patterns in how error categories distribute across iterations (Fig.~\ref{fig:error_evolution}):

\begin{enumerate}
    \item \textbf{All three major error categories co-occur from the first repair iteration.} At iteration~1, LB errors account for 42.3\% of active problems (compilable code that fails test cases), MS errors for 38.5\% (ownership and borrow-checker violations), and SL errors for 19.2\% (syntax and type mismatches). 
    This early co-occurrence of diverse error types motivates per-category temperature differentiation in ESTS rather than a one-size-fits-all repair strategy.
    \item \textbf{MS and SL errors decrease as repair progresses.} MS errors decline steadily as the RAG module injects ownership-reasoning templates ($\mathcal{T}_{MS}$) at moderate entropy ($\theta_{MS} = 0.5$). These templates guide the LLM away from \texttt{unsafe} escape hatches.
    SL errors resolve efficiently when they appear, aided by low-entropy deterministic repair ($\theta_{SL} = 0.1$) and syntax-correction templates ($\mathcal{T}_{SL}$).
    \item \textbf{LB errors are the most persistent and increasingly dominant.} LB errors constitute 42.3\% of active problems at iteration~1 and rise to 85.7\% by iteration~19. 
    The root causes of these errors, such as algorithmic flaws, off-by-one errors, and incorrect data structure choices, are decoupled from syntax and ownership. These errors constitute the primary bottleneck for unsolved problems. The ESTS module activates high-entropy exploration ($\theta_{LB} = 1.2$) to escape local optima by generating structurally diverse alternatives. This mechanism functions as a mutation operator in the search-based software engineering sense~\cite{sbse}.
\end{enumerate}

These patterns represent statistical tendencies across the dataset, not strict sequential phases within individual trajectories. 
A single trajectory may revisit earlier error categories after a stagnation reset or high-temperature exploration, as demonstrated by the C\_422\_4 case study (Section~\ref{subsec:case_study}), whose repair path cycles through 
MS\,$\to$\,\allowbreak AF\,$\to$\,\allowbreak LB$\times$2\,$\to$\,\allowbreak MS$\times$3\,$\to$\,\allowbreak RESET$_1$\,$\to$\,\allowbreak SL$\times$3\,$\to$\,\allowbreak RESET$_2$\,$\to$\,\allowbreak SUCCESS
(where each RESET fires at the fourth consecutive same-category occurrence, i.e., $\kappa_{local} = 4 > N_L$, and the $\times$3 counts the stagnation iterations preceding the triggering one).

\subsubsection{Role of ESTS in Convergence}
The ablation study (Section~\ref{subsec:rq3}) confirms that removing ESTS causes the largest single-component solve-rate drop ($-$11.86 pp). 
The convergence perspective reveals why. 
Without error categorization, the repair cycle oscillates between equivalent error states because uniform-temperature sampling fails to match the exploration intensity required by different error categories. 
The Local-Loop stagnation escape ($N_L = 3$) further aids convergence by triggering a context reset when the same category repeats. 
This reset functions as a random restart that prevents budget exhaustion on saturated trajectories (see sensitivity analysis in Section~\ref{subsec:threshold_sensitivity}). 
Across the 104-problem dataset, RESET events fire in 16 distinct problems (31 events total). Of these, 14 (87.5\%) are ultimately solved. 
This high recovery rate indicates that the stagnation escape mechanism effectively redirects the search toward productive trajectories.

\subsubsection{Case Study: Iterative Repair of Problem C\_422\_4}\label{subsec:case_study}

As a concrete illustration of the error-driven repair dynamics, we trace the full transformation of Problem~4 from Weekly Contest~422. 
The C source implements a \texttt{count\allowbreak Balanced\allowbreak Permutations} function that counts the distinct permutations of a digit string such that the sum of digits at even-indexed positions equals the sum at odd-indexed positions, computed modulo $10^9 + 7$. 
The solution employs a memoized recursive search over a three-dimensional state space (digit value, remaining sum, remaining count) with precomputed binomial coefficients via Pascal's triangle.

\textbf{Baseline comparison on this problem.}
This problem is exceptionally difficult for zero-shot LLM transformation. 
Across 200 independent samples, not a single transformation passes the test suite. The pass@1 through pass@100 estimates (following Chen et al.~\cite{codex}) are all 0.0\%, despite 42.0\% of samples compiling successfully. 
The zero-shot LLM can produce syntactically plausible Rust code for this problem but consistently fails to preserve the precise modular arithmetic and combinatorial logic required for functional correctness. 
AdaTrans solves this problem in 14 iterations (one initial generation plus 13 repair steps). The zero-shot approach fails on all 200 samples, yet the iterative repair cycle converges to a correct solution.
The following trace details how the framework converges through error-driven adaptation. 
In the iteration headers below, the category label (e.g., MS, LB) denotes the ESTS error classification that drove the repair attempt, derived from the diagnostic signal of the preceding iteration, and may differ from the compilation outcome of the current attempt.

\textbf{Iteration 0 (INIT, $\theta = 1.0$).}
The initial zero-shot transformation mirrors the C global-variable pattern using Rust's \texttt{static mut} declarations, which require \texttt{unsafe} blocks for every access:
\begin{lstlisting}[language=Rust, frame=none, numbers=none, basicstyle=\ttfamily\footnotesize]
static mut CNT: [usize; MAX_DIGITS] = [0; MAX_DIGITS];
static mut DP: [[[Option<i64>; MAX_COUNT]; MAX_SUM];
    MAX_DIGITS] = [[[None; MAX_COUNT]; ...];
\end{lstlisting}
The compiler reports \texttt{E0308} and \texttt{E0277}: ``cannot multiply \texttt{i64} by \texttt{usize}''---a type mismatch between the \texttt{i64} accumulator and the \texttt{usize} binomial coefficient array. This is classified as a Memory-Semantic (MS) error because it originates from the interaction between Rust's strict numeric type system and the ownership of mutable global state.

\textbf{Iterations 1--4 (MS$\to$AF$\to$LB$\to$LB).}
The ESTS module routes the MS error to $\theta_{MS} = 0.5$, and the RAG module retrieves a type-repair template from $\mathcal{T}_{MS}$. 
Iteration~1 fails with the same type mismatch. 
At iteration~2, the error shifts to an Ambiguous-Fallback (AF) classification ($\theta = 1.0$), and the code compiles but panics at runtime with ``attempt to subtract with overflow'' in the recursive DP function. 
Iteration~3 produces a wrong answer (expected output 2, actual output 0), classified as Logic-Behavioral (LB) at $\theta_{LB} = 1.2$. 
Iteration~4, still driven by the LB classification at $\theta_{LB} = 1.2$, regresses to a compile error: the high-temperature exploration restructures the numeric types from \texttt{i64} to \texttt{isize} and introduces new \texttt{Mul<usize>} trait violations.

\textbf{Iterations 5--7 (MS$\times$3, stagnation).}
Iterations~5--7 repeat the same MS-classified \texttt{E0308}/\texttt{E0277} compile error: ``cannot multiply \texttt{isize} by \texttt{usize}.'' 
Despite RAG-guided ownership repair at $\theta_{MS} = 0.5$, the model cycles through equivalent type-conversion attempts without resolving the underlying mismatch between signed and unsigned integer types in the DP recurrence.

\textbf{Iteration 8 (RESET$_1$, $\theta = 1.0$).}
Once the MS category persists beyond three consecutive repetitions ($\kappa_{local} = 4 > N_L = 3$), the Local-Loop stagnation escape triggers a context reset. 
The system discards the accumulated repair context and re-derives a fresh Rust transformation from the C source at baseline temperature $T = 1.0$. 
The regenerated code retains \texttt{static mut} globals but shifts all arrays to \texttt{i64}, partially resolving the earlier type conflicts. A residual \texttt{E0308} remains (``you can convert a \texttt{usize} to an \texttt{i64}'').

\textbf{Iterations 9--11 (SL$\times$3, stagnation).}
Iterations~9--11 attempt to fix the remaining type mismatch, now classified as Syntactic-Linking (SL) at $\theta_{SL} = 0.1$.
Low-temperature repairs produce minimal edits that fail to resolve the function signature incompatibility and stagnate at the same error.

\textbf{Iteration 12 (RESET$_2$, $\theta = 1.0$).}
A second RESET fires once the SL category persists beyond three consecutive repetitions ($\kappa_{local} = 4 > N_L$). 
The fresh regeneration produces a restructured solution: \texttt{static mut} globals are eliminated in favor of function parameters and heap-allocated \texttt{Vec} containers, which removes all \texttt{unsafe} blocks. However, a new \texttt{E0308} error arises: ``expected \texttt{\&[usize; 10]}, found \texttt{\&Vec<usize>}''---a fixed-size vs.\ dynamic array mismatch in the recursive function signature.

\textbf{Iteration 13 (SUCCESS).}
The final iteration resolves the array type mismatch by unifying all container types as \texttt{Vec<usize>} and restructuring the DP function to accept slice references. The solution passes all test cases with zero \texttt{unsafe} blocks:
\begin{lstlisting}[language=Rust, frame=none, numbers=none, basicstyle=\ttfamily\footnotesize]
fn dfs(i: usize, s: usize, c: usize,
       left_s: &[usize], left_c: &[usize],
       cnt: &[usize], cb: &[Vec<usize>],
       dp: &mut Vec<Vec<Vec<Option<usize>>>>,
       r1: &[usize]) -> usize { ... }
\end{lstlisting}
The transformed code replaces C-style global mutable state and pointer arithmetic with Rust-idiomatic parameter passing, \texttt{Vec}-based dynamic allocation, and \texttt{u128} intermediate arithmetic for modular multiplication.

\textbf{Key observations from this case study.}
(1)~The problem is beyond zero-shot capability (pass@100 = 0.0\% across 200 samples), yet AdaTrans solves it in 14 iterations through adaptive error-driven repair. 
(2)~All four error categories manifest in the repair trajectory (MS, AF, LB, SL), which exercises the full spectrum of adaptive temperature scheduling in ESTS. 
(3)~Two distinct RESET events occur: the first escapes MS-category stagnation (iterations~5--7), and the second escapes SL-category stagnation (iterations~9--11). This confirms that the Local-Loop mechanism handles diverse error types. 
(4)~The second RESET (iteration~12) induces a qualitative architectural shift, namely a move from \texttt{static mut} globals with pervasive \texttt{unsafe} blocks to safe, parameterized functions. This shift illustrates that stagnation escape can trigger structural improvements beyond local repair. 
(5)~The final solution eliminates all \texttt{unsafe} code and preserves the modular arithmetic correctness of the original C implementation. This outcome shows that both safety and functional equivalence are achievable on this combinatorially complex algorithm.

\textbf{Comparative case: c2rust transpilation failures.}
Two further cases highlight the limitations of traditional AST-level transpilation. In both, c2rust fails to produce valid Rust code, while AdaTrans generates safe, idiomatic alternatives.

Problem 3 from Weekly Contest 418 (\texttt{constructGridLayout}) uses variable-length array (VLA) initializers, a C99 feature with no direct Rust equivalent:
\begin{lstlisting}[language=C, frame=none, numbers=none, basicstyle=\ttfamily\footnotesize]
int son[n][2] = {}, sou[n] = {};
\end{lstlisting}
c2rust cannot represent this construct in Rust and omits the core \texttt{constructGridLayout} function entirely. It generates only the \texttt{main} wrapper with raw \texttt{libc} pointers and \texttt{unsafe} blocks. 
AdaTrans, by contrast, transforms the VLA into idiomatic Rust vectors:
\begin{lstlisting}[language=Rust, frame=none, numbers=none, basicstyle=\ttfamily\footnotesize]
let mut son: Vec<Vec<usize>> = vec![vec![0; 2]; n];
let mut sou: Vec<usize> = vec![0; n];
\end{lstlisting}
The resulting code uses safe slicing, standard I/O (\texttt{std::io}), and no \texttt{unsafe} blocks, while preserving the algorithm's correctness.

Problem 3 from Weekly Contest 430 (\texttt{numberOfSubsequences}) defines multiple helper functions (\texttt{hash}, \texttt{hash\_insert}, \texttt{hash\_get}, \texttt{gcd}) as nested functions inside the main function, a GCC C extension not supported by standard C99. c2rust reports ``function definition is not allowed here'' errors for each nested function and produces structurally invalid output. 
AdaTrans successfully refactors the nested functions into top-level Rust functions and replaces the manual hash table with Rust's \texttt{HashMap}:
\begin{lstlisting}[language=Rust, frame=none, numbers=none, basicstyle=\ttfamily\footnotesize]
fn gcd(mut a: i32, mut b: i32) -> i32 {
    while b > 0 { let temp = b; b = a % b; a = temp; }
    a
}
fn number_of_subsequences(nums: &[i32]) -> i64 {
    let mut hash_table = HashMap::new();
    // ...
}
\end{lstlisting}
These cases illustrate a key advantage of LLM-based transformation over deterministic AST rewriting. 
The LLM can infer semantic intent from C idioms and re-express it with Rust-native abstractions, whereas the rigid syntax-level mapping of c2rust fails when the source code relies on constructs that have no direct syntactic counterpart in Rust.

%% file: sections/threats.tex
\section{Threats to Validity}\label{sec:threats}

We organize potential threats to validity into external threats (factors that limit the generalizability of our findings) and internal threats (factors that may affect the correctness of the experimental methodology).

\subsection{External Validity}

\textbf{Dataset representativeness.}
Our evaluation uses 104 algorithmic problems from LeetCode Weekly Contests (Contests 413--438). 
These problems are self-contained, single-file programs with standard I/O interfaces. 
This scope may not fully represent the challenges of systems programming or large-scale codebases with complex pointer arithmetic and multi-file dependencies. 
We mitigate selection bias by sampling problems across 26 distinct contest sessions that cover diverse algorithmic paradigms, including dynamic programming, graph traversal, and combinatorial optimization. 
The file-level focus is deliberate, as it targets incremental migration, where Rust is gradually adopted within legacy C environments~\cite{rustforlinux}.

\textbf{Baseline tool adaptation.}
EvoC2Rust~\cite{evotrans}, Tymcrat~\cite{tymcrat}, and PtrTrans~\cite{ptrtrans} were designed for project-level or function-level transformation. 
We adapted each tool for our file-level, single-file evaluation protocol. 
This adaptation may disadvantage these tools relative to their intended use cases. 
We mitigated this concern by following the documentation of each tool, resolving only the minimum set of integration issues required to run on our dataset, and applying the same compilation and I/O consistency protocol to all methods.

\textbf{LLM selection.}
All experiments employ \texttt{gpt-4o-mini} as the underlying LLM. The modular design of AdaTrans is model-agnostic. 
However, performance may vary across LLM backends. 
We selected \texttt{gpt-4o-mini} for its balance of cost efficiency and capability, which enabled the large number of experimental runs required for ablation and sensitivity analyses. 
Commercial LLM providers may also silently update model weights behind a fixed API endpoint, which can introduce run-to-run variance beyond sampling stochasticity. Pinning the dated snapshot (\texttt{gpt-4o-mini-2024-07-18}) mitigates but cannot fully eliminate this risk. 
Future work should evaluate AdaTrans with other LLMs to assess transferability.

\textbf{Data contamination.}
LLM-based transformation results may be inflated if evaluation problems overlap with the model's pretraining data. 
We mitigate this risk by sourcing problems exclusively from LeetCode Weekly Contests released after September~2024, which postdates the knowledge cutoff (October~2023) of \texttt{gpt-4o-mini} (Section~\ref{subsec:dataset}). 
While temporal separation cannot guarantee zero overlap, it provides a principled safeguard against memorization of solution patterns.

\subsection{Internal Validity}

\textbf{Evaluation oracle.}
We adopt I/O consistency over 10--200 fuzzing-generated test cases per problem (100 for the majority) as the operational definition of functional equivalence. This choice follows established practice in the code transformation literature~\cite{codex,evotrans}. 
While this test suite covers both typical and boundary-condition inputs, it cannot guarantee equivalence on untested inputs. 
Our normalization operator $\text{Norm}(\cdot)$ mitigates benign formatting discrepancies through whitespace collapsing. All outputs in our dataset are exact integers or strings, so only whitespace normalization is applied.

\textbf{Non-determinism.}
LLMs are inherently stochastic~\cite{self_consistency}. 
We quantify run-to-run variance by executing AdaTrans three times on the full 104-problem dataset. The solve rates are 84.62\%, 79.81\%, and 78.85\%. These runs yield a mean of 81.09\% $\pm$ 3.09\% (standard deviation). 
The moderate variance indicates stable performance across runs. The core components of AdaTrans, namely the deterministic RAG retrieval function $f_{retrieve}$ and the ESTS mapping $\mathcal{F}_{ESTS}$, reduce run-to-run variance compared to unconstrained LLM generation. The zero-shot LLM baseline mitigates non-determinism through pass@$k$ estimation over 200 independent samples~\cite{codex}.

\textbf{Temperature parameter sensitivity.}
The ESTS module relies on three temperature parameters ($\theta_{SL}$, $\theta_{MS}$, $\theta_{LB}$) whose default values are selected based on the entropy-rigidity mapping. We do not claim these values are globally optimal. 
A diagnostic sensitivity analysis on the full 104-problem dataset (Section~\ref{subsec:temp_sensitivity}) suggests that the framework appears reasonably stable under moderate perturbations within each entropy tier and that the directional mapping (low for syntax, moderate for ownership, high for logic) is a consistent factor in effectiveness.

%% file: sections/related-work.tex
\section{Related Work}\label{sec:related}

\subsection{C-to-Rust Transformation Tools}

Rule-based tools such as c2rust~\cite{c2rust} transform C source code through AST-level rewriting. 
The output compiles but relies heavily on \texttt{unsafe} constructs. Extensive manual refactoring is therefore required to achieve safe Rust.
Ling et al.~\cite{10.1145/3510454.3528640} proposed CRustS, which applies source-to-source transformation rules expressed in the TXL language~\cite{CORDY2006190} to reduce the scope of unsafe expressions in c2rust output.
Emre et al.~\cite{laet} proposed Laertes, which lifts a subset of raw pointers in c2rust output into safe Rust references by using the borrow checker as a lifetime oracle.
Emre et al.~\cite{10.1145/3586046} later examined the limits of this strategy and found that imprecision in the Rust safety checker labels many safe pointer manipulations as potentially unsafe, which constrains the fraction of pointers that can be lifted into safe references.
More recently, LLM-based approaches have emerged.
Wang et al.~\cite{evotrans} introduced EvoC2Rust, which refines LLM-generated Rust through an evolutionary, skeleton-guided strategy. 
Hong and Ryu~\cite{tymcrat} proposed Tymcrat, an LLM-based type-migration tool that replaces C types with appropriate Rust types and iteratively fixes the resulting type errors using compiler feedback.
Yuan et al.~\cite{ptrtrans} proposed PtrTrans, a framework that constructs a C-Rust Pointer Knowledge Graph to encode pointer ownership and lifetime information.
Luo et al.~\cite{11186085} proposed IRENE, which integrates rule-augmented retrieval, structured code summarization, and error-driven repair to improve the safety and semantic consistency of LLM-based C-to-Rust translation.
Cai et al.~\cite{10.1007/978-3-032-00828-2_16} proposed RustMap, a project-scale approach that decomposes a C program into dependency-guided units and uses input--output equivalence checks together with compiler and test feedback to refine the translated Rust code.
These approaches operate primarily at the project or function level. AdaTrans targets the complementary file-level scenario.

\subsection{LLM-Based Code Generation and Transformation}

The Codex model~\cite{codex} demonstrated that LLMs trained on large code corpora can solve non-trivial programming tasks and established the basis for tools such as GitHub Copilot. 
Code Llama~\cite{codellama} and GPT-4~\cite{gpt4} further advanced the state of the art in code generation and instruction following. 
Jiang et al.~\cite{llm_code_survey} provide a survey of LLM applications in code generation. For cross-language transformation, Rozi\`{e}re et al.~\cite{transcoder} proposed an unsupervised approach based on back-transformation that does not require parallel corpora. 
Pre-trained code models such as CodeBERT~\cite{codebert}, PLBART~\cite{plbart}, InCoder~\cite{incoder}, CodeT5~\cite{wang2021codet5}, and GrammarT5~\cite{10.1145/3597503.3639125} provide strong code representations, yet they lack the compilation feedback needed to reliably produce safe, compilable target code.

\subsection{Neural and Cross-Language Code Translation}

Cross-language code translation has been studied extensively beyond the C-to-Rust setting.
Building on unsupervised neural machine translation~\cite{transcoder}, Liu et al.~\cite{10.1109/ICSE48619.2023.00072} proposed SDA-Trans, which incorporates syntax structure and domain knowledge to improve cross-lingual transfer, particularly for languages unseen during pre-training.
Shi et al.~\cite{shi-etal-2022-natural} introduced execution-result-based minimum Bayes risk decoding, which selects among candidate programs by approximating semantic equivalence through execution on a small number of test inputs.
Rule-based and customizable transpilers offer an alternative to purely neural translation.
Wang et al.~\cite{10.1145/3586034} proposed DuoGlot, which incrementally constructs user-guided transpilation rules to translate Python to JavaScript with high accuracy.
Doeraene~\cite{item_4964f148bf9841a49a57b06963268153} presented Scala.js, a type-directed interoperability framework between statically typed Scala and dynamically typed JavaScript.
Sun et al.~\cite{10.1145/3510003.3510140} applied context-aware code translation to the related task of code search by translating code snippets into natural language descriptions.

Because automatically translated code frequently contains residual errors, several works focus on human collaboration and error localization.
Weisz et al.~\cite{10.1145/3397481.3450656} studied how software engineers tolerate imperfect AI-generated translations and identified interface features that aid error detection.
Liu et al.~\cite{10.1109/TSE.2024.3379583} proposed hmCodeTrans, an interactive human-machine method that feeds engineer edits back to the model for retranslation.
Malyala et al.~\cite{10.1109/ICSE-NIER58687.2023.00017} analyzed common failure patterns of unsupervised translators and combined rule-based pre- and post-processing with a neural model to improve translation.
Wang et al.~\cite{10.1145/3611643.3616322} proposed TransMap, which pinpoints the location of semantic mistakes in neural code translation to reduce debugging effort.

The evaluation of code translation has also received dedicated attention.
Jiao et al.~\cite{10.1109/ASE56229.2023.00114} developed a taxonomy of translation tasks by complexity and knowledge dependence and showed that existing benchmarks are biased toward trivial token-level translation.
Macedo et al.~\cite{10.1145/3650105.3652301} found that the output format of LLM-generated translations substantially affects execution-based evaluation and proposed extraction methods to recover source code from model responses.
Pendyala and Thakur~\cite{PENDYALA2026100811} presented an automated evaluation and explainability framework that combines compilation and execution validation with token-level feature attribution across multiple code LLMs.
These studies inform our use of an execution-based oracle and exact output matching for evaluating functional equivalence (Section~\ref{sec:setup}).

\subsection{Retrieval-Augmented Generation for Code}

Lewis et al.~\cite{rag} introduced the RAG framework, which combines a pre-trained retriever with a generative model to produce more factual and contextually grounded outputs. 
In the code domain, Xia et al.~\cite{coderepair_llm} demonstrated that large pre-trained language models substantially improve automated program repair, especially when fix-template information is incorporated. 
AdaTrans builds on this foundation with a retrieval module that maps compiler error signatures to error-specific repair templates and official Rust documentation through a Rust Violation-Fix Graph.

\subsection{Adaptive Strategies for LLM Generation}

Wang et al.~\cite{self_consistency} introduced self-consistency decoding, which samples multiple reasoning paths and selects the most consistent answer. 
Zhu et al.~\cite{temp_sensitivity} showed that adaptive temperature control yields substantial gains over fixed-temperature sampling in code generation. 
Harman et al.~\cite{sbse} survey search-based optimization for software engineering and review how metaheuristic search can explore large solution spaces.
Xuan et al.~\cite{DBLP:journals/tse/XuanJRL12} applied a backbone-based multilevel metaheuristic to the large-scale next release problem, illustrating how search scales to large software optimization spaces.
AdaTrans draws on these perspectives by treating temperature selection as a feedback-driven search problem.

\subsection{Automated Program Repair}

Automated program repair (APR) generates patches for buggy programs by using test suites as correctness specifications. 
Xuan et al.~\cite{nopol} proposed Nopol, which encodes conditional statement repair as a Satisfiability Modulo Theory (SMT) problem and synthesizes patches from runtime traces. 
Xiong et al.~\cite{acs} introduced ACS, which improves repair precision through condition synthesis guided by code context and document analysis.
Effective repair often depends on accurate fault localization.
Xuan and Monperrus~\cite{DBLP:conf/sigsoft/XuanM14} proposed test case purification, which decomposes failing tests to sharpen spectrum-based fault localization.
Gu et al.~\cite{GuXZZFXQ19} predicted whether a crashing fault resides within the stack trace to guide crash localization, and Liu et al.~\cite{https://doi.org/10.1002/smr.70099} studied nested exceptions in Java crash reports to support crash reproduction.
At the project level, Xuan et al.~\cite{DBLP:journals/tkde/XuanJHRZLW15} applied data reduction techniques to improve the efficiency of bug triage.
The iterative repair cycle of AdaTrans shares the generate-and-validate paradigm with APR. Compiler diagnostics and test feedback serve as the correctness specification, and the ESTS module adapts the repair strategy based on error category, analogous to how APR techniques select repair templates based on fault localization.

\subsection{Memory Safety in Rust}

The Rust ownership and borrowing system provides compile-time memory safety guarantees without garbage collection~\cite{rust_study_1}~\cite{rust_study_2}. 
The formal soundness of this system has been verified through the RustBelt project~\cite{rust_borrow_checker}. 
Qin et al.~\cite{unsafe_rust} studied how and why programmers write \texttt{unsafe} code in real-world Rust projects and found that \texttt{unsafe} is commonly used to support low-level control that safe Rust does not permit.
Cui et al.~\cite{unsafe_migration} further examined safety requirements across unsafe API boundaries and identified a set of safety properties that
programmers must satisfy.
These findings highlight the challenge facing any C-to-Rust tool. The generated code must satisfy the Rust ownership rules rather than circumvent them.

%% file: sections/conclusion.tex
\section{Conclusion}\label{sec:conclusion}

This paper presents AdaTrans, a framework for automated C-to-Rust code transformation that aligns LLM-generated code with the ownership and memory safety guarantees of Rust. 
AdaTrans integrates three mechanisms. 
First, a Strategy-Driven RAG mechanism maps compiler error signatures to targeted repair strategies through a Rust Violation-Fix Graph. 
Second, an Error-Stratified Transformation Strategy (ESTS) classifies compiler diagnostics into semantic error categories and adjusts the repair temperature per category to balance exploitation for syntax repairs with exploration for algorithmic mutations. 
Third, a multi-stage validation pipeline enforces both compilability and functional equivalence through a closed-loop generate-verify-repair cycle.

Our evaluation on 104 C algorithm problems demonstrates that AdaTrans achieves a mean compilation pass rate of 95.51\% ($\pm$ 1.11\%) and a mean solve rate of 81.09\% ($\pm$ 3.09\%) across three independent runs, validated against a fuzz oracle of 10--200 test cases per problem, with a mean \texttt{unsafe} file rate of 1.19\%.
AdaTrans surpasses the pass@100 estimate (70.58\%), a high-budget reference point for brute-force independent sampling under the same backbone, through a single iterative repair trajectory of at most 20 iterations. 
Ablation studies confirm the individual contributions of both the RAG and ESTS components. Removing ESTS reduces the mean solve rate to 69.23\%, removing RAG to 71.15\%, and removing both to 46.79\%.

Several directions remain for future investigation. 
First, the current file-level transformation scope should be extended to support project-level migration, where multi-file dependencies, build system integration, and cross-module type sharing introduce additional complexities. 
Second, evaluating AdaTrans with stronger and more diverse LLM backends (beyond \texttt{gpt-4o-mini}) would provide deeper insights into the model-agnostic properties of the framework and its scalability to frontier models. 
Third, expanding the error taxonomy and RAG knowledge base to cover a broader range of C constructs, including POSIX system calls, hardware-specific intrinsics, and third-party library bindings, would enhance applicability to real-world systems programming scenarios. 
Finally, integrating formal verification techniques alongside the current testing-based validation could provide stronger guarantees of semantic equivalence and move beyond I/O consistency toward provable correctness.

%% file: sections/declarations.tex
\section*{Declarations}

\subsection*{Conflict of Interest}
The authors declare that they have no competing interests.

\subsection*{Funding}
This work was supported by [FUNDING INFORMATION REMOVED FOR REVIEW].

\subsection*{Data Availability}
All data, code, and experimental artifacts are available in the replication package (see Section~\ref{sec:setup}).

\subsection*{Author Contributions}
Xiaofan Liu conceived and designed the study, performed the experiments, and wrote the original draft. 
Zecan Li assisted with data collection. 
Zhuang Zhao participated in the discussion of the ideas and the revision of the manuscript. 
Ziqi Shuai assisted with the experimental infrastructure. 
Qi Xin participated in the discussion of the ideas and the revision of the manuscript. 
Jifeng Xuan supervised the project and contributed to the review and editing of the paper. 
All authors read and approved the final manuscript.

%% file: sections/acknowledgments.tex
\section*{Acknowledgments}


%% file: references.bib
@misc{c2rust,
  author={{Immunant, Inc.} and {Galois, Inc.}},
  title={{C2Rust}: Migrating {C} Code to {Rust}},
  year={2018},
  howpublished={\url{https://github.com/immunant/c2rust}},
  note={Open-source transpiler. Accessed: 2025}
}

@article{laet,
    author = {Emre, Mehmet and Schroeder, Ryan and Dewey, Kyle and Hardekopf, Ben},
    title = {Translating C to safer Rust},
    year = {2021},
    issue_date = {October 2021},
    publisher = {Association for Computing Machinery},
    address = {New York, NY, USA},
    volume = {5},
    number = {OOPSLA},
    url = {https://doi.org/10.1145/3485498},
    doi = {10.1145/3485498},
    journal = {Proc. ACM Program. Lang.},
    month = oct,
    articleno = {121},
    numpages = {29}
}

@misc{evotrans,
  author       = {Chaofan Wang and
                  Tingrui Yu and
                  Chen Xie and
                  Jie Wang and
                  Dong Chen and
                  Wenrui Zhang and
                  Yuling Shi and
                  Xiaodong Gu and
                  Beijun Shen},
  title        = {EvoC2Rust: A Skeleton-guided Framework for Project-Level {C}-to-{Rust} Translation},
  booktitle    = {Proceedings of the IEEE/ACM 48th International Conference on Software Engineering: Software Engineering in Practice (ICSE-SEIP)},
  year         = {2026},
  publisher    = {IEEE/ACM},
  numpages     = {12},
  url          = {https://arxiv.org/abs/2508.04295}
}

@article{ptrtrans,
  author       = {Zhiqiang Yuan and
                  Wenjun Mao and
                  Zhuo Chen and
                  Xiyue Shang and
                  Chong Wang and
                  Yiling Lou and
                  Xin Peng},
  title        = {Project-Level {C}-to-{Rust} Translation via Synergistic Integration of Knowledge Graphs and Large Language Models},
  journal      = {Proceedings of the ACM on Software Engineering},
  volume       = {3},
  number       = {FSE},
  articleno    = {162},
  numpages     = {24},
  year         = {2026},
  publisher    = {ACM},
  address      = {Montreal, QC, Canada},
  month        = jul,
  note         = {Conference: ACM SIGSOFT International Symposium on the Foundations of Software Engineering (FSE 2026)},
}

@article{tymcrat,
  author       = {Jaemin Hong and
                  Sukyoung Ryu},
  title        = {Type-migrating C-to-Rust translation using a large language model},
  journal      = {Empir. Softw. Eng.},
  volume       = {30},
  number       = {1},
  pages        = {3},
  year         = {2025},
  url          = {https://doi.org/10.1007/s10664-024-10573-2},
  doi          = {10.1007/S10664-024-10573-2},
  bibsource    = {dblp computer science bibliography, https://dblp.org}
}

@misc{codex,
      title={Evaluating Large Language Models Trained on Code}, 
      author={Mark Chen and Jerry Tworek and Heewoo Jun and Qiming Yuan and Henrique Ponde de Oliveira Pinto and Jared Kaplan and Harri Edwards and Yuri Burda and Nicholas Joseph and Greg Brockman and Alex Ray and Raul Puri and Gretchen Krueger and Michael Petrov and Heidy Khlaaf and Girish Sastry and Pamela Mishkin and Brooke Chan and Scott Gray and Nick Ryder and Mikhail Pavlov and Alethea Power and Lukasz Kaiser and Mohammad Bavarian and Clemens Winter and Philippe Tillet and Felipe Petroski Such and Dave Cummings and Matthias Plappert and Fotios Chantzis and Elizabeth Barnes and Ariel Herbert-Voss and William Hebgen Guss and Alex Nichol and Alex Paino and Nikolas Tezak and Jie Tang and Igor Babuschkin and Suchir Balaji and Shantanu Jain and William Saunders and Christopher Hesse and Andrew N. Carr and Jan Leike and Josh Achiam and Vedant Misra and Evan Morikawa and Alec Radford and Matthew Knight and Miles Brundage and Mira Murati and Katie Mayer and Peter Welinder and Bob McGrew and Dario Amodei and Sam McCandlish and Ilya Sutskever and Wojciech Zaremba},
      year={2021},
      eprint={2107.03374},
      archivePrefix={arXiv},
      primaryClass={cs.LG},
      url={https://arxiv.org/abs/2107.03374}, 
}

@misc{codellama,
      title={Code Llama: Open Foundation Models for Code}, 
      author={Baptiste Rozière and Jonas Gehring and Fabian Gloeckle and Sten Sootla and Itai Gat and Xiaoqing Ellen Tan and Yossi Adi and Jingyu Liu and Romain Sauvestre and Tal Remez and Jérémy Rapin and Artyom Kozhevnikov and Ivan Evtimov and Joanna Bitton and Manish Bhatt and Cristian Canton Ferrer and Aaron Grattafiori and Wenhan Xiong and Alexandre Défossez and Jade Copet and Faisal Azhar and Hugo Touvron and Louis Martin and Nicolas Usunier and Thomas Scialom and Gabriel Synnaeve},
      year={2024},
      eprint={2308.12950},
      archivePrefix={arXiv},
      primaryClass={cs.CL},
      url={https://arxiv.org/abs/2308.12950}, 
}

@misc{gpt4,
      title={GPT-4 Technical Report}, 
      author={OpenAI and Josh Achiam and Steven Adler and Sandhini Agarwal and Lama Ahmad and Ilge Akkaya and Florencia Leoni Aleman and Diogo Almeida and Janko Altenschmidt and Sam Altman and Shyamal Anadkat and Red Avila and Igor Babuschkin and Suchir Balaji and Valerie Balcom and Paul Baltescu and Haiming Bao and Mohammad Bavarian and Jeff Belgum and Irwan Bello and Jake Berdine and Gabriel Bernadett-Shapiro and Christopher Berner and Lenny Bogdonoff and Oleg Boiko and Madelaine Boyd and Anna-Luisa Brakman and Greg Brockman and Tim Brooks and Miles Brundage and Kevin Button and Trevor Cai and Rosie Campbell and Andrew Cann and Brittany Carey and Chelsea Carlson and Rory Carmichael and Brooke Chan and Che Chang and Fotis Chantzis and Derek Chen and Sully Chen and Ruby Chen and Jason Chen and Mark Chen and Ben Chess and Chester Cho and Casey Chu and Hyung Won Chung and Dave Cummings and Jeremiah Currier and Yunxing Dai and Cory Decareaux and Thomas Degry and Noah Deutsch and Damien Deville and Arka Dhar and David Dohan and Steve Dowling and Sheila Dunning and Adrien Ecoffet and Atty Eleti and Tyna Eloundou and David Farhi and Liam Fedus and Niko Felix and Simón Posada Fishman and Juston Forte and Isabella Fulford and Leo Gao and Elie Georges and Christian Gibson and Vik Goel and Tarun Gogineni and Gabriel Goh and Rapha Gontijo-Lopes and Jonathan Gordon and Morgan Grafstein and Scott Gray and Ryan Greene and Joshua Gross and Shixiang Shane Gu and Yufei Guo and Chris Hallacy and Jesse Han and Jeff Harris and Yuchen He and Mike Heaton and Johannes Heidecke and Chris Hesse and Alan Hickey and Wade Hickey and Peter Hoeschele and Brandon Houghton and Kenny Hsu and Shengli Hu and Xin Hu and Joost Huizinga and Shantanu Jain and Shawn Jain and Joanne Jang and Angela Jiang and Roger Jiang and Haozhun Jin and Denny Jin and Shino Jomoto and Billie Jonn and Heewoo Jun and Tomer Kaftan and Łukasz Kaiser and Ali Kamali and Ingmar Kanitscheider and Nitish Shirish Keskar and Tabarak Khan and Logan Kilpatrick and Jong Wook Kim and Christina Kim and Yongjik Kim and Jan Hendrik Kirchner and Jamie Kiros and Matt Knight and Daniel Kokotajlo and Łukasz Kondraciuk and Andrew Kondrich and Aris Konstantinidis and Kyle Kosic and Gretchen Krueger and Vishal Kuo and Michael Lampe and Ikai Lan and Teddy Lee and Jan Leike and Jade Leung and Daniel Levy and Chak Ming Li and Rachel Lim and Molly Lin and Stephanie Lin and Mateusz Litwin and Theresa Lopez and Ryan Lowe and Patricia Lue and Anna Makanju and Kim Malfacini and Sam Manning and Todor Markov and Yaniv Markovski and Bianca Martin and Katie Mayer and Andrew Mayne and Bob McGrew and Scott Mayer McKinney and Christine McLeavey and Paul McMillan and Jake McNeil and David Medina and Aalok Mehta and Jacob Menick and Luke Metz and Andrey Mishchenko and Pamela Mishkin and Vinnie Monaco and Evan Morikawa and Daniel Mossing and Tong Mu and Mira Murati and Oleg Murk and David Mély and Ashvin Nair and Reiichiro Nakano and Rajeev Nayak and Arvind Neelakantan and Richard Ngo and Hyeonwoo Noh and Long Ouyang and Cullen O'Keefe and Jakub Pachocki and Alex Paino and Joe Palermo and Ashley Pantuliano and Giambattista Parascandolo and Joel Parish and Emy Parparita and Alex Passos and Mikhail Pavlov and Andrew Peng and Adam Perelman and Filipe de Avila Belbute Peres and Michael Petrov and Henrique Ponde de Oliveira Pinto and Michael and Pokorny and Michelle Pokrass and Vitchyr H. Pong and Tolly Powell and Alethea Power and Boris Power and Elizabeth Proehl and Raul Puri and Alec Radford and Jack Rae and Aditya Ramesh and Cameron Raymond and Francis Real and Kendra Rimbach and Carl Ross and Bob Rotsted and Henri Roussez and Nick Ryder and Mario Saltarelli and Ted Sanders and Shibani Santurkar and Girish Sastry and Heather Schmidt and David Schnurr and John Schulman and Daniel Selsam and Kyla Sheppard and Toki Sherbakov and Jessica Shieh and Sarah Shoker and Pranav Shyam and Szymon Sidor and Eric Sigler and Maddie Simens and Jordan Sitkin and Katarina Slama and Ian Sohl and Benjamin Sokolowsky and Yang Song and Natalie Staudacher and Felipe Petroski Such and Natalie Summers and Ilya Sutskever and Jie Tang and Nikolas Tezak and Madeleine B. Thompson and Phil Tillet and Amin Tootoonchian and Elizabeth Tseng and Preston Tuggle and Nick Turley and Jerry Tworek and Juan Felipe Cerón Uribe and Andrea Vallone and Arun Vijayvergiya and Chelsea Voss and Carroll Wainwright and Justin Jay Wang and Alvin Wang and Ben Wang and Jonathan Ward and Jason Wei and CJ Weinmann and Akila Welihinda and Peter Welinder and Jiayi Weng and Lilian Weng and Matt Wiethoff and Dave Willner and Clemens Winter and Samuel Wolrich and Hannah Wong and Lauren Workman and Sherwin Wu and Jeff Wu and Michael Wu and Kai Xiao and Tao Xu and Sarah Yoo and Kevin Yu and Qiming Yuan and Wojciech Zaremba and Rowan Zellers and Chong Zhang and Marvin Zhang and Shengjia Zhao and Tianhao Zheng and Juntang Zhuang and William Zhuk and Barret Zoph},
      year={2024},
      eprint={2303.08774},
      archivePrefix={arXiv},
      primaryClass={cs.CL},
      url={https://arxiv.org/abs/2303.08774}, 
}

@inproceedings{coderepair_llm,
author = {Xia, Chunqiu Steven and Wei, Yuxiang and Zhang, Lingming},
title = {Automated Program Repair in the Era of Large Pre-Trained Language Models},
year = {2023},
isbn = {9781665457019},
publisher = {IEEE Press},
address = {Piscataway, NJ, USA},
url = {https://doi.org/10.1109/ICSE48619.2023.00129},
doi = {10.1109/ICSE48619.2023.00129},
booktitle = {Proceedings of the 45th International Conference on Software Engineering},
pages = {1482–1494},
numpages = {13},
location = {Melbourne, Victoria, Australia},
series = {ICSE '23}
}

@inproceedings{rag,
author = {Lewis, Patrick and Perez, Ethan and Piktus, Aleksandra and Petroni, Fabio and Karpukhin, Vladimir and Goyal, Naman and K\"{u}ttler, Heinrich and Lewis, Mike and Yih, Wen-tau and Rockt\"{a}schel, Tim and Riedel, Sebastian and Kiela, Douwe},
title = {Retrieval-augmented generation for knowledge-intensive NLP tasks},
year = {2020},
isbn = {9781713829546},
publisher = {Curran Associates Inc.},
address = {Red Hook, NY, USA},
booktitle = {Proceedings of the 34th International Conference on Neural Information Processing Systems},
articleno = {793},
numpages = {16},
location = {Vancouver, BC, Canada},
series = {NIPS '20}
}

@inproceedings{self_consistency,
  author       = {Xuezhi Wang and
                  Jason Wei and
                  Dale Schuurmans and
                  Quoc V. Le and
                  Ed H. Chi and
                  Sharan Narang and
                  Aakanksha Chowdhery and
                  Denny Zhou},
  title        = {Self-Consistency Improves Chain of Thought Reasoning in Language Models},
  booktitle    = {The Eleventh International Conference on Learning Representations,
                  {ICLR} 2023, Kigali, Rwanda, May 1-5, 2023},
  publisher    = {OpenReview.net},
  address = {Amherst, MA, USA},
  numpages     = {24},
  year         = {2023},
  url          = {https://openreview.net/forum?id=1PL1NIMMrw},
  bibsource    = {dblp computer science bibliography, https://dblp.org}
}

@inproceedings{transcoder,
author = {Roziere, Baptiste and Lachaux, Marie-Anne and Chanussot, Lowik and Lample, Guillaume},
title = {Unsupervised translation of programming languages},
year = {2020},
isbn = {9781713829546},
publisher = {Curran Associates Inc.},
address = {Red Hook, NY, USA},
booktitle = {Proceedings of the 34th International Conference on Neural Information Processing Systems},
articleno = {1730},
numpages = {11},
location = {Vancouver, BC, Canada},
series = {NIPS '20}
}

@inproceedings{incoder,
  author       = {Daniel Fried and
                  Armen Aghajanyan and
                  Jessy Lin and
                  Sida Wang and
                  Eric Wallace and
                  Freda Shi and
                  Ruiqi Zhong and
                  Scott Yih and
                  Luke Zettlemoyer and
                  Mike Lewis},
  title        = {InCoder: {A} Generative Model for Code Infilling and Synthesis},
  booktitle    = {The Eleventh International Conference on Learning Representations,
                  {ICLR} 2023, Kigali, Rwanda, May 1-5, 2023},
  publisher    = {OpenReview.net},
  address      = {Kigali, Rwanda},
  year         = {2023},
  numpages     = {26}, 
  url          = {https://openreview.net/forum?id=hQwb-lbM6EL},
  bibsource    = {dblp computer science bibliography, https://dblp.org}
}

@inproceedings{unsafe_rust,
author = {Qin, Boqin and Chen, Yilun and Yu, Zeming and Song, Linhai and Zhang, Yiying},
title = {Understanding memory and thread safety practices and issues in real-world Rust programs},
year = {2020},
isbn = {9781450376136},
publisher = {Association for Computing Machinery},
address = {New York, NY, USA},
url = {https://doi.org/10.1145/3385412.3386036},
doi = {10.1145/3385412.3386036},
booktitle = {Proceedings of the 41st ACM SIGPLAN Conference on Programming Language Design and Implementation},
pages = {763–779},
numpages = {17},
location = {London, UK},
series = {PLDI 2020}
}

@inproceedings{rust_study_1,
author = {Matsakis, Nicholas D. and Klock, Felix S.},
title = {The rust language},
year = {2014},
isbn = {9781450332170},
publisher = {Association for Computing Machinery},
address = {New York, NY, USA},
url = {https://doi.org/10.1145/2663171.2663188},
doi = {10.1145/2663171.2663188},
booktitle = {Proceedings of the 2014 ACM SIGAda Annual Conference on High Integrity Language Technology},
pages = {103–104},
numpages = {2},
location = {Portland, Oregon, USA},
series = {HILT '14}
}

@article{rust_study_2,
author = {Matsakis, Nicholas D. and Klock, Felix S.},
title = {The rust language},
year = {2014},
issue_date = {December 2014},
publisher = {Association for Computing Machinery},
address = {New York, NY, USA},
volume = {34},
number = {3},
issn = {1094-3641},
url = {https://doi.org/10.1145/2692956.2663188},
doi = {10.1145/2692956.2663188},
journal = {Ada Lett.},
month = oct,
pages = {103–104},
numpages = {2}
}

@article{llm_code_survey,
  author       = {Juyong Jiang and
                  Fan Wang and
                  Jiasi Shen and
                  Sungju Kim and
                  Sung Hun Kim},
  title        = {A Survey on Large Language Models for Code Generation},
  journal      = {{ACM} Trans. Softw. Eng. Methodol.},
  volume       = {35},
  number       = {2},
  pages        = {58:1--58:72},
  year         = {2026},
  url          = {https://doi.org/10.1145/3747588},
  doi          = {10.1145/3747588},
  bibsource    = {dblp computer science bibliography, https://dblp.org}
}

@inproceedings{ctorust_userstudy,
  author       = {Ruishi Li and
                  Bo Wang and
                  Tianyu Li and
                  Prateek Saxena and
                  Ashish Kundu},
  title        = {Translating {C} To Rust: Lessons from a User Study},
  booktitle    = {32nd Annual Network and Distributed System Security Symposium, {NDSS}
                  2025, San Diego, California, USA, February 24-28, 2025},
  publisher    = {The Internet Society},
  address   = {Reston, VA, USA},
  year         = {2025},
  numpages  = {18},
  url          = {https://www.ndss-symposium.org/ndss-paper/translating-c-to-rust-lessons-from-a-user-study/},
  bibsource    = {dblp computer science bibliography, https://dblp.org}
}

@inproceedings{temp_sensitivity,
  author       = {Yuqi Zhu and
                  Jia Li and
                  Ge Li and
                  Yunfei Zhao and
                  Jia Li and
                  Zhi Jin and
                  Hong Mei},
  editor       = {Michael J. Wooldridge and
                  Jennifer G. Dy and
                  Sriraam Natarajan},
  title        = {Hot or Cold? Adaptive Temperature Sampling for Code Generation with
                  Large Language Models},
  booktitle    = {Thirty-Eighth {AAAI} Conference on Artificial Intelligence, {AAAI}
                  2024, Thirty-Sixth Conference on Innovative Applications of Artificial
                  Intelligence, {IAAI} 2024, Fourteenth Symposium on Educational Advances
                  in Artificial Intelligence, {EAAI} 2014, February 20-27, 2024, Vancouver,
                  Canada},
  pages        = {437--445},
  publisher    = {{AAAI} Press},
  address = {Palo Alto, CA, USA},
  year         = {2024},
  url          = {https://doi.org/10.1609/aaai.v38i1.27798},
  doi          = {10.1609/AAAI.V38I1.27798},
  bibsource    = {dblp computer science bibliography, https://dblp.org}
}

@article{sbse,
author = {Harman, Mark and Mansouri, S. Afshin and Zhang, Yuanyuan},
title = {Search-based software engineering: Trends, techniques and applications},
year = {2012},
issue_date = {November 2012},
publisher = {Association for Computing Machinery},
address = {New York, NY, USA},
volume = {45},
number = {1},
issn = {0360-0300},
url = {https://doi.org/10.1145/2379776.2379787},
doi = {10.1145/2379776.2379787},
journal = {ACM Comput. Surv.},
month = dec,
articleno = {11},
numpages = {61}
}

@article{rust_borrow_checker,
author = {Jung, Ralf and Jourdan, Jacques-Henri and Krebbers, Robbert and Dreyer, Derek},
title = {RustBelt: securing the foundations of the Rust programming language},
year = {2017},
issue_date = {January 2018},
publisher = {Association for Computing Machinery},
address = {New York, NY, USA},
volume = {2},
number = {POPL},
url = {https://doi.org/10.1145/3158154},
doi = {10.1145/3158154},
journal = {Proc. ACM Program. Lang.},
month = dec,
articleno = {66},
numpages = {34}
}

@inproceedings{plbart,
    title = "Unified Pre-training for Program Understanding and Generation",
    author = "Ahmad, Wasi  and
      Chakraborty, Saikat  and
      Ray, Baishakhi  and
      Chang, Kai-Wei",
    editor = "Toutanova, Kristina  and
      Rumshisky, Anna  and
      Zettlemoyer, Luke  and
      Hakkani-Tur, Dilek  and
      Beltagy, Iz  and
      Bethard, Steven  and
      Cotterell, Ryan  and
      Chakraborty, Tanmoy  and
      Zhou, Yichao",
    booktitle = "Proceedings of the 2021 Conference of the North American Chapter of the Association for Computational Linguistics: Human Language Technologies",
    month = jun,
    year = "2021",
    address = "Online",
    publisher = "Association for Computational Linguistics",
    url = "https://aclanthology.org/2021.naacl-main.211/",
    doi = "10.18653/v1/2021.naacl-main.211",
    pages = "2655--2668"
}

@inproceedings{codebert,
    title = "{C}ode{BERT}: A Pre-Trained Model for Programming and Natural Languages",
    author = "Feng, Zhangyin  and
      Guo, Daya  and
      Tang, Duyu  and
      Duan, Nan  and
      Feng, Xiaocheng  and
      Gong, Ming  and
      Shou, Linjun  and
      Qin, Bing  and
      Liu, Ting  and
      Jiang, Daxin  and
      Zhou, Ming",
    editor = "Cohn, Trevor  and
      He, Yulan  and
      Liu, Yang",
    booktitle = "Findings of the Association for Computational Linguistics: EMNLP 2020",
    month = nov,
    year = "2020",
    address = "Online",
    publisher = "Association for Computational Linguistics",
    url = "https://aclanthology.org/2020.findings-emnlp.139/",
    doi = "10.18653/v1/2020.findings-emnlp.139",
    pages = "1536--1547"
}

@inproceedings{unsafe_migration,
author = {Cui, Mohan and Sun, Shuran and Xu, Hui and Zhou, Yangfan},
title = {Is unsafe an Achilles' Heel? A Comprehensive Study of Safety Requirements in Unsafe Rust Programming},
year = {2024},
isbn = {9798400702174},
publisher = {Association for Computing Machinery},
address = {New York, NY, USA},
url = {https://doi.org/10.1145/3597503.3639136},
doi = {10.1145/3597503.3639136},
booktitle = {Proceedings of the IEEE/ACM 46th International Conference on Software Engineering},
articleno = {106},
numpages = {13},
location = {Lisbon, Portugal},
series = {ICSE '24}
}

@inproceedings{rustforlinux,
author = {Li, Hongyu and Guo, Liwei and Yang, Yexuan and Wang, Shangguang and Xu, Mengwei},
title = {An empirical study of rust-for-Linux: the success, dissatisfaction, and compromise},
year = {2024},
isbn = {978-1-939133-41-0},
publisher = {USENIX Association},
address = {USA},
booktitle = {Proceedings of the 2024 USENIX Conference on Usenix Annual Technical Conference},
articleno = {27},
numpages = {19},
location = {Santa Clara, CA, USA},
series = {USENIX ATC'24}
}

@article{difftesting,
  title={Differential Testing for Software},
  author={McKeeman, William M.},
  journal={Digital Technical Journal},
  volume={10},
  number={1},
  pages={100--107},
  year={1998}
}

@inproceedings{acs,
author = {Xiong, Yingfei and Wang, Jie and Yan, Runfa and Zhang, Jiachen and Han, Shi and Huang, Gang and Zhang, Lu},
title = {Precise condition synthesis for program repair},
year = {2017},
isbn = {9781538638682},
publisher = {IEEE Press},
address = {Piscataway, NJ, USA},
url = {https://doi.org/10.1109/ICSE.2017.45},
doi = {10.1109/ICSE.2017.45},
booktitle = {Proceedings of the 39th International Conference on Software Engineering},
pages = {416–426},
numpages = {11},
location = {Buenos Aires, Argentina},
series = {ICSE '17}
}

@article{nopol,
  author={Xuan, Jifeng and Martinez, Matias and DeMarco, Favio and Clément, Maxime and Marcote, Sebastian Lamelas and Durieux, Thomas and Le Berre, Daniel and Monperrus, Martin},
  journal={IEEE Transactions on Software Engineering}, 
  title={Nopol: Automatic Repair of Conditional Statement Bugs in Java Programs}, 
  year={2017},
  volume={43},
  number={1},
  pages={34-55},
  doi={10.1109/TSE.2016.2560811}
}

@article{csmith_a,
author = {Yang, Xuejun and Chen, Yang and Eide, Eric and Regehr, John},
title = {Finding and understanding bugs in C compilers},
year = {2011},
issue_date = {June 2011},
publisher = {Association for Computing Machinery},
address = {New York, NY, USA},
volume = {46},
number = {6},
issn = {0362-1340},
url = {https://doi.org/10.1145/1993316.1993532},
doi = {10.1145/1993316.1993532},
journal = {SIGPLAN Not.},
month = jun,
pages = {283–294},
numpages = {12}
}

@inproceedings{csmith_b,
author = {Yang, Xuejun and Chen, Yang and Eide, Eric and Regehr, John},
title = {Finding and understanding bugs in C compilers},
year = {2011},
isbn = {9781450306638},
publisher = {Association for Computing Machinery},
address = {New York, NY, USA},
url = {https://doi.org/10.1145/1993498.1993532},
doi = {10.1145/1993498.1993532},
booktitle = {Proceedings of the 32nd ACM SIGPLAN Conference on Programming Language Design and Implementation},
pages = {283–294},
numpages = {12},
location = {San Jose, California, USA},
series = {PLDI '11}
}

@misc{adatrans_replication,
  author={Liu, Xiaofan and Li, Zecan and Zhao, Zhuang and Shuai, Ziqi and Xuan, Jifeng},
  title={{AdaTrans} Replication Package},
  year={2025},
  howpublished={\url{https://github.com/SlainTroyard/adatrans_dev}},
  note={Source code, prompts, knowledge base, and experimental data.}
}

@inproceedings{wang2021codet5,
    title = "{C}ode{T}5: Identifier-aware Unified Pre-trained Encoder-Decoder Models for Code Understanding and Generation",
    author = "Wang, Yue  and
      Wang, Weishi  and
      Joty, Shafiq  and
      Hoi, Steven C.H.",
    editor = "Moens, Marie-Francine  and
      Huang, Xuanjing  and
      Specia, Lucia  and
      Yih, Scott Wen-tau",
    booktitle = "Proceedings of the 2021 Conference on Empirical Methods in Natural Language Processing",
    month = nov,
    year = "2021",
    address = "Online and Punta Cana, Dominican Republic",
    publisher = "Association for Computational Linguistics",
    url = "https://aclanthology.org/2021.emnlp-main.685/",
    doi = "10.18653/v1/2021.emnlp-main.685",
    pages = "8696--8708"
}

@inproceedings{10.1145/3510003.3510140,
author = {Sun, Weisong and Fang, Chunrong and Chen, Yuchen and Tao, Guanhong and Han, Tingxu and Zhang, Quanjun},
title = {Code search based on context-aware code translation},
year = {2022},
isbn = {9781450392211},
publisher = {Association for Computing Machinery},
address = {New York, NY, USA},
url = {https://doi.org/10.1145/3510003.3510140},
doi = {10.1145/3510003.3510140},
booktitle = {Proceedings of the 44th International Conference on Software Engineering},
pages = {388–400},
numpages = {13},
location = {Pittsburgh, Pennsylvania},
series = {ICSE '22}
}

@inproceedings{10.1145/3650105.3652301,
author = {Macedo, Marcos and Tian, Yuan and Cogo, Filipe and Adams, Bram},
title = {Exploring the Impact of the Output Format on the Evaluation of Large Language Models for Code Translation},
year = {2024},
isbn = {9798400706097},
publisher = {Association for Computing Machinery},
address = {New York, NY, USA},
url = {https://doi.org/10.1145/3650105.3652301},
doi = {10.1145/3650105.3652301},
booktitle = {Proceedings of the 2024 IEEE/ACM First International Conference on AI Foundation Models and Software Engineering},
pages = {57–68},
numpages = {12},
location = {Lisbon, Portugal},
series = {FORGE '24}
}

@inproceedings{10.1145/3597503.3639125,
author = {Zhu, Qihao and Liang, Qingyuan and Sun, Zeyu and Xiong, Yingfei and Zhang, Lu and Cheng, Shengyu},
title = {GrammarT5: Grammar-Integrated Pretrained Encoder-Decoder Neural Model for Code},
year = {2024},
isbn = {9798400702174},
publisher = {Association for Computing Machinery},
address = {New York, NY, USA},
url = {https://doi.org/10.1145/3597503.3639125},
doi = {10.1145/3597503.3639125},
booktitle = {Proceedings of the IEEE/ACM 46th International Conference on Software Engineering},
articleno = {76},
numpages = {13},
location = {Lisbon, Portugal},
series = {ICSE '24}
}

@article{10.1109/TSE.2024.3379583,
author = {Liu, Jiaqi and Zhang, Fengming and Zhang, Xin and Yu, Zhiwen and Wang, Liang and Zhang, Yao and Guo, Bin},
title = {hmCodeTrans: Human–Machine Interactive Code Translation},
year = {2024},
issue_date = {May 2024},
publisher = {IEEE Press},
volume = {50},
number = {5},
issn = {0098-5589},
url = {https://doi.org/10.1109/TSE.2024.3379583},
doi = {10.1109/TSE.2024.3379583},
journal = {IEEE Trans. Softw. Eng.},
month = may,
pages = {1163–1181},
numpages = {19}
}

@inproceedings{10.1109/ICSE-NIER58687.2023.00017,
author = {Malyala, Aniketh and Zhou, Katelyn and Ray, Baishakhi and Chakraborty, Saikat},
title = {On ML-Based Program Translation: Perils and Promises},
year = {2023},
isbn = {9798350300390},
publisher = {IEEE Press},
address = {Piscataway, NJ, USA},
url = {https://doi.org/10.1109/ICSE-NIER58687.2023.00017},
doi = {10.1109/ICSE-NIER58687.2023.00017},
booktitle = {Proceedings of the 45th International Conference on Software Engineering: New Ideas and Emerging Results},
pages = {60–65},
numpages = {6},
location = {Melbourne, Australia},
series = {ICSE-NIER '23}
}

@inproceedings{shi-etal-2022-natural,
    title = "Natural Language to Code Translation with Execution",
    author = "Shi, Freda  and
      Fried, Daniel  and
      Ghazvininejad, Marjan  and
      Zettlemoyer, Luke  and
      Wang, Sida I.",
    editor = "Goldberg, Yoav  and
      Kozareva, Zornitsa  and
      Zhang, Yue",
    booktitle = "Proceedings of the 2022 Conference on Empirical Methods in Natural Language Processing",
    month = dec,
    year = "2022",
    address = "Abu Dhabi, United Arab Emirates",
    publisher = "Association for Computational Linguistics",
    url = "https://aclanthology.org/2022.emnlp-main.231/",
    doi = "10.18653/v1/2022.emnlp-main.231",
    pages = "3533--3546"
}

@inproceedings{10.1109/ASE56229.2023.00114,
author = {Jiao, Mingsheng and Yu, Tingrui and Li, Xuan and Qiu, Guanjie and Gu, Xiaodong and Shen, Beijun},
title = {On the Evaluation of Neural Code Translation: Taxonomy and Benchmark},
year = {2024},
isbn = {9798350329964},
publisher = {IEEE Press},
address = {Piscataway, NJ, USA},
url = {https://doi.org/10.1109/ASE56229.2023.00114},
doi = {10.1109/ASE56229.2023.00114},
booktitle = {Proceedings of the 38th IEEE/ACM International Conference on Automated Software Engineering},
pages = {1529–1541},
numpages = {13},
location = {Echternach, Luxembourg},
series = {ASE '23}
}

@inproceedings{10.1145/3397481.3450656,
author = {Weisz, Justin D. and Muller, Michael and Houde, Stephanie and Richards, John and Ross, Steven I. and Martinez, Fernando and Agarwal, Mayank and Talamadupula, Kartik},
title = {Perfection Not Required? Human-AI Partnerships in Code Translation},
year = {2021},
isbn = {9781450380171},
publisher = {Association for Computing Machinery},
address = {New York, NY, USA},
url = {https://doi.org/10.1145/3397481.3450656},
doi = {10.1145/3397481.3450656},
booktitle = {Proceedings of the 26th International Conference on Intelligent User Interfaces},
pages = {402–412},
numpages = {11},
location = {College Station, TX, USA},
series = {IUI '21}
}

@article{PENDYALA2026100811,
title = {Rosetta-XAI: An automated evaluation and explainability framework for code translation models},
journal = {Software Impacts},
volume = {27},
pages = {100811},
year = {2026},
issn = {2665-9638},
doi = {https://doi.org/10.1016/j.simpa.2026.100811},
url = {https://www.sciencedirect.com/science/article/pii/S2665963826000011},
author = {Vishnu S. Pendyala and Neha Bais Thakur}
}

@inproceedings{10.1109/ICSE48619.2023.00072,
author = {Liu, Fang and Li, Jia and Zhang, Li},
title = {Syntax and Domain Aware Model for Unsupervised Program Translation},
year = {2023},
isbn = {9781665457019},
publisher = {IEEE Press},
address = {Piscataway, NJ, USA},
url = {https://doi.org/10.1109/ICSE48619.2023.00072},
doi = {10.1109/ICSE48619.2023.00072},
booktitle = {Proceedings of the 45th International Conference on Software Engineering},
pages = {755–767},
numpages = {13},
location = {Melbourne, Victoria, Australia},
series = {ICSE '23}
}

@article{CORDY2006190,
title = {The TXL source transformation language},
journal = {Science of Computer Programming},
volume = {61},
number = {3},
pages = {190-210},
year = {2006},
note = {Special Issue on The Fourth Workshop on Language Descriptions, Tools, and Applications (LDTA ’04)},
issn = {0167-6423},
doi = {https://doi.org/10.1016/j.scico.2006.04.002},
url = {https://www.sciencedirect.com/science/article/pii/S0167642306000669},
author = {James R. Cordy}
}

@inproceedings{10.1145/3611643.3616322,
author = {Wang, Bo and Li, Ruishi and Li, Mingkai and Saxena, Prateek},
title = {TransMap: Pinpointing Mistakes in Neural Code Translation},
year = {2023},
isbn = {9798400703270},
publisher = {Association for Computing Machinery},
address = {New York, NY, USA},
url = {https://doi.org/10.1145/3611643.3616322},
doi = {10.1145/3611643.3616322},
booktitle = {Proceedings of the 31st ACM Joint European Software Engineering Conference and Symposium on the Foundations of Software Engineering},
pages = {999–1011},
numpages = {13},
location = {San Francisco, CA, USA},
series = {ESEC/FSE 2023}
}

@article{10.1145/3586046,
author = {Emre, Mehmet and Boyland, Peter and Parekh, Aesha and Schroeder, Ryan and Dewey, Kyle and Hardekopf, Ben},
title = {Aliasing Limits on Translating C to Safe Rust},
year = {2023},
issue_date = {April 2023},
publisher = {Association for Computing Machinery},
address = {New York, NY, USA},
volume = {7},
number = {OOPSLA1},
url = {https://doi.org/10.1145/3586046},
doi = {10.1145/3586046},
journal = {Proc. ACM Program. Lang.},
month = apr,
articleno = {94},
numpages = {29}
}

@article{10.1145/3586034,
author = {Wang, Bo and Kolluri, Aashish and Nikoli\'{c}, Ivica and Baluta, Teodora and Saxena, Prateek},
title = {User-Customizable Transpilation of Scripting Languages},
year = {2023},
issue_date = {April 2023},
publisher = {Association for Computing Machinery},
address = {New York, NY, USA},
volume = {7},
number = {OOPSLA1},
url = {https://doi.org/10.1145/3586034},
doi = {10.1145/3586034},
journal = {Proc. ACM Program. Lang.},
month = apr,
articleno = {82},
numpages = {29}
}

@inproceedings{10.1145/3510454.3528640,
author = {Ling, Michael and Yu, Yijun and Wu, Haitao and Wang, Yuan and Cordy, James R. and Hassan, Ahmed E.},
title = {In rust we trust: a transpiler from unsafe C to safer rust},
year = {2022},
isbn = {9781450392235},
publisher = {Association for Computing Machinery},
address = {New York, NY, USA},
url = {https://doi.org/10.1145/3510454.3528640},
doi = {10.1145/3510454.3528640},
booktitle = {Proceedings of the ACM/IEEE 44th International Conference on Software Engineering: Companion Proceedings},
pages = {354–355},
numpages = {2},
location = {Pittsburgh, Pennsylvania},
series = {ICSE '22}
}

@INPROCEEDINGS {11186085,
author = { Luo, Feng and Ji, Kexing and Gao, Cuiyun and Gao, Shuzheng and Feng, Jia and Liu, Kui and Xia, Xin and Lyu, Michael R. },
booktitle = { 2025 IEEE International Conference on Software Maintenance and Evolution (ICSME) },
title = {{ Integrating Rules and Semantics for LLM-Based C-to-Rust Translation }},
year = {2025},
volume = {},
ISSN = {},
pages = {685-696},
doi = {10.1109/ICSME64153.2025.00069},
url = {https://doi.ieeecomputersociety.org/10.1109/ICSME64153.2025.00069},
publisher = {IEEE Computer Society},
address = {Los Alamitos, CA, USA},
month =sep}

@misc{item_4964f148bf9841a49a57b06963268153, title={Scala.js: Type-Directed Interoperability with Dynamically Typed Languages}, url={https://infoscience.epfl.ch/handle/20.500.14299/97425}, abstractNote={Interoperability between statically typed and dynamically typed languages is increasingly important, as can be witnessed by the many statically typed languages targeting JavaScript. Interoperating with both the object-oriented and functional features of JavaScript is essential, if only to manipulate the DOM, yet existing languages have very poor support for this. We present Scala.js, a dialect of Scala compiling to JavaScript. Its interoperability system is based on a powerful and intuitive framework for type-directed interoperability with dynamically typed languages. The framework combines facade types for JavaScript values; user-defined, implicit, type-directed cross-language conversions; and a Dynamic type building on facade types and implicit conversions. It accommodates both the functional and object-oriented features of Scala and JavaScript, and provides very natural interoperability between the two languages. It is expressive enough to represent the DOM and jQuery APIs, among others, both in its statically typed and dynamically typed flavors.}, author={Doeraene, Sébastien}, year={2013} }

@inproceedings{10.1007/978-3-032-00828-2_16,
author = {Cai, Xuemeng and Liu, Jiakun and Huang, Xiping and Yu, Yijun and Wu, Haitao and Li, Chunmiao and Wang, Bo and Yusuf, Imam Nur Bani and Jiang, Lingxiao},
title = {RustMap: Towards Project-Scale C-to-Rust Migration via Program Analysis and LLM},
year = {2025},
isbn = {978-3-032-00827-5},
publisher = {Springer-Verlag},
address = {Berlin, Heidelberg},
url = {https://doi.org/10.1007/978-3-032-00828-2_16},
doi = {10.1007/978-3-032-00828-2_16},
booktitle = {Engineering of Complex Computer Systems: 29th International Conference, ICECCS 2025, Hangzhou, China, July 2–4, 2025, Proceedings},
pages = {283–302},
numpages = {20},
location = {Hangzhou, China}
}

@article{DBLP:journals/tkde/XuanJHRZLW15,
  author    = {Jifeng Xuan and
               He Jiang and
               Yan Hu and
               Zhilei Ren and
               Weiqin Zou and
               Zhongxuan Luo and
               Xindong Wu},
  title     = {Towards Effective Bug Triage with Software Data Reduction Techniques},
  journal   = {{IEEE} Trans. Knowl. Data Eng.},
  volume    = {27},
  number    = {1},
  pages     = {264--280},
  year      = {2015},
  url       = {https://doi.org/10.1109/TKDE.2014.2324590},
  doi       = {10.1109/TKDE.2014.2324590},
  bibsource = {dblp computer science bibliography, https://dblp.org}
}

@article{DBLP:journals/tse/XuanJRL12,
  author    = {Jifeng Xuan and
               He Jiang and
               Zhilei Ren and
               Zhongxuan Luo},
  title     = {Solving the Large Scale Next Release Problem with a Backbone-Based Multilevel Algorithm},
  journal   = {{IEEE} Trans. Software Eng.},
  volume    = {38},
  number    = {5},
  pages     = {1195--1212},
  year      = {2012},
  url       = {https://doi.org/10.1109/TSE.2011.92},
  doi       = {10.1109/TSE.2011.92},
  bibsource = {dblp computer science bibliography, https://dblp.org}
}

@inproceedings{DBLP:conf/sigsoft/XuanM14,
author = {Xuan, Jifeng and Monperrus, Martin},
title = {Test case purification for improving fault localization},
year = {2014},
isbn = {9781450330565},
publisher = {Association for Computing Machinery},
address = {New York, NY, USA},
url = {https://doi.org/10.1145/2635868.2635906},
doi = {10.1145/2635868.2635906},
booktitle = {Proceedings of the 22nd ACM SIGSOFT International Symposium on Foundations of Software Engineering},
pages = {52–63},
numpages = {12},
location = {Hong Kong, China},
series = {FSE 2014}
}

@article{GuXZZFXQ19,
  author    = {Yongfeng Gu and
               Jifeng Xuan and
               Hongyu Zhang and
               Lanxin Zhang and
               Qingna Fan and
               Xiaoyuan Xie and
               Tieyun Qian},
  title     = {Does the fault reside in a stack trace? Assisting crash localization by predicting crashing fault residence},
  journal   = {Journal of Systems and Software},
  volume    = {148},
  pages     = {88--104},
  year      = {2019},
  url       = {https://doi.org/10.1016/j.jss.2018.11.004},
  doi       = {10.1016/j.jss.2018.11.004},
  bibsource = {dblp computer science bibliography, https://dblp.org}
}

@article{https://doi.org/10.1002/smr.70099,
author = {Liu, Shaoting and Xu, Haiyan and Xin, Qi and Xuan, Jifeng},
title = {Exceptions Can Be Nested: An Exploratory Study on Nested Exceptions in Java Crashes},
journal = {Journal of Software: Evolution and Process},
volume = {38},
number = {4},
pages = {e70099},
doi = {https://doi.org/10.1002/smr.70099},
url = {https://onlinelibrary.wiley.com/doi/abs/10.1002/smr.70099},
eprint = {https://onlinelibrary.wiley.com/doi/pdf/10.1002/smr.70099},
year = {2026}
}

@misc{openai2024gpt4ominidocs,
  author       = {{OpenAI}},
  title        = {Models - GPT-4o-mini},
  year         = {2024},
  howpublished = {\url{https://platform.openai.com/docs/models/gpt-4o-mini}},
  note         = {Accessed: 2026-06-08}
}
